\documentclass[a4paper,11pt]{article}
\pdfoutput=1 

\usepackage{jheppub} 
                     
\usepackage{graphicx} 
\usepackage{epsfig}
\usepackage{epsf}
\usepackage{epstopdf}
\usepackage{hyperref}

\usepackage{amsmath}  
\usepackage{pdfsync}
\usepackage{slashed}

\newcommand{\bea}{\begin{eqnarray}} 
\newcommand{\eea}{\end{eqnarray}}

\def\absp{|\bp|}
\def\abspr{|\bp'|}

\newcommand{\symb}{\textrm{symb}}
\newcommand{\symbi}{\symb^{-1}}

\newcommand{\ps}{\slashed{p}}

\newcommand{\pipe}{|}

\usepackage[T1]{fontenc} 

\def\t #1{\tau_{#1}}
\def\non{\nonumber\\}

\def\e{\,{\rm e}}

\def\veps#1{\varepsilon_{#1}}

\def\multilin{\Big\vert_{\veps_1\veps_2\cdots \veps_N}}

\def\half{{1\over 2}}

\def\fourth{{1\over4}}

\def\Eins{{\mathchoice {\rm 1\mskip-4mu l} {\rm 1\mskip-4mu l}
{\rm 1\mskip-4.5mu l} {\rm 1\mskip-5mu l}}}
\def\Z{{\mathchoice {\hbox{$\sf\textstyle Z\kern-0.4em Z$}}
{\hbox{$\sf\textstyle Z\kern-0.4em Z$}}
{\hbox{$\sf\scriptstyle Z\kern-0.3em Z$}}
{\hbox{$\sf\scriptscriptstyle Z\kern-0.2em Z$}}}}
\def\abs#1{\left| #1\right|}

                \def\slash#1{\slashed{#1}}

\newcommand{\slD}{\,\raise.15ex\hbox{$/$}\kern-.27em\hbox{$\!\!\!D$}}
\newcommand{\slpartial}{\raise.15ex\hbox{$/$}\kern-.57em\hbox{$\partial$}}

\def\epseps#1#2{\varepsilon_{#1}\cdot\varepsilon_{#2}}

\def\no{\noindent}

\def\kinq{{1\over 4}\dot q^2}

\def\be{\begin{equation}}
\def\ee{\end{equation}\noindent}
\def\bear{\begin{eqnarray}}
\def\ear{\end{eqnarray}\noindent}
\def\bec{\blue\begin{equation}}
\def\eec{\end{equation}\black\noindent}
\def\bearc{\blue\begin{eqnarray}}
\def\earc{\end{eqnarray}\black\noindent}
\def\benn{\begin{enumerate}}
\def\enn{\end{enumerate}}
\def\ee{&=&}

\def\ab{{\alpha\beta}}

\def\mn{{\mu\nu}}

\def\tr{{\rm tr}\,}

\def\e{\,{\rm e}}

\def\bp{{\bf p}}

\def\b0{{\bf 0}}

\def\4piTD{{(4\pi T)}^{-{D\over 2}}}
\def\4piT4{{(4\pi T)}^{-2}}

\newcommand{\s}{\slashed}
\def\veps{\varepsilon}

\newcommand{\epsilonilon}{\varepsilon}
\def\veps{\epsilonilon}
\newcommand{\bone}{1\!\!1}

\newcommand{\jhat}[1]{\hspace{0.3em}\widehat{\hspace{-0.4em}#1\hspace{-0.4em}}\hspace{0.4em}}

\title{\boldmath Worldline master formulas for the dressed electron propagator, part 2: On-shell amplitudes}

\author[a]{N. Ahmadiniaz,}
\author[f]{V. M. Banda Guzm\'an,}
\author[c,d]{F. Bastianelli,}
\author[e,d]{O. Corradini,}
\author[f]{J.P. Edwards}
\author[f]{and C. Schubert}

\affiliation[a]{ Helmholtz-Zentrum Dresden-Rossendorf, Bautzner Landstra\ss e 400, 01328 Dresden, Germany}
\affiliation[c]{Dipartimento di Fisica e Astronomia ``Augusto Righi'', Universit\`a di Bologna, Via Irnerio 46, I-40126 Bologna, Italy}
\affiliation[d]{INFN, Sezione di Bologna, Via Irnerio 46, I-40126 Bologna, Italy}
\affiliation[e]{Dipartimento di Scienze Fisiche, Informatiche e Matematiche,
 Universit\`a degli Studi di Modena e Reggio Emilia, Via Campi 213/A, I-41125 Modena, Italy}
\affiliation[f]{Instituto de F\'isica y Matem\'aticas
Universidad Michoacana de San Nicol\'as de Hidalgo
Edificio C-3, Apdo. Postal 2-82
C.P. 58040, Morelia, Michoac\'an, M\'exico}
\emailAdd{n.ahmadiniaz@hzdr.de}
\emailAdd{victor.banda@umich.mx}
\emailAdd{bastianelli@bo.infn.it} 
\emailAdd{olindo.corradini@unimore.it}
\emailAdd{jedwards@ifm.umich.mx}
\emailAdd{schubert@ifm.umich.mx}

\abstract{
In the first part of this series, we employed the second-order formalism and the ``symbol'' map to construct a particle path-integral representation of the electron propagator in a background electromagnetic field, suitable for open fermion-line calculations. 
Its main advantages are the avoidance of long products of Dirac matrices, 
and its ability to unify whole sets of Feynman diagrams related by permutation of photon legs along the fermion lines. 
We obtained a Bern-Kosower type master formula for the fermion propagator, dressed with $N$ photons, in terms of the ``$N$-photon kernel,'' where this kernel appears also in ``subleading'' terms involving only $N-1$ of the $N$ photons.

In this sequel, we focus on the application of the formalism to the calculation of on-shell amplitudes and cross sections. 
Universal formulas are obtained for the fully polarised matrix elements of the fermion propagator dressed with an arbitrary
number of photons, as well as for the corresponding spin-averaged cross sections. 
A major simplification of the on-shell case is that the subleading terms drop out, but we also pinpoint other, less obvious simplifications. 

We use integration by parts to achieve manifest transversality of these amplitudes at the integrand level and exploit this property using the spinor helicity technique.
We give a simple proof of the vanishing of the matrix element for ``all $+$'' photon helicities in the massless case,  
and find a novel relation between the scalar and spinor spin-averaged cross sections in the massive case. 
Testing the formalism on the standard linear Compton scattering process, we find that it reproduces the known results with remarkable efficiency. Further applications and generalisations are pointed out. 
}

\begin{document} 
\maketitle
\flushbottom

\section{Introduction}
\label{sec:intro}
In the first part of this series \cite{130}, simply to be called `I' in the following, we developed a novel path-integral representation
of the electron propagator in an external electromagnetic field suitable for practical calculations in the worldline approach to
QED \cite{feynman:pr80,feynman:pr84,polyakovbook,strassler1,strassler2,5,15,41, UsRep}, in particular for tree-level amplitudes involving multiple photons and related quantities. 
In contrast to the standard first-order Dirac formalism, it is based on the less-known but equivalent second order fermion approach
to spinor QED \cite{feygel,hostler,berdun,morgan}. Let us briefly retrace the main steps of this derivation, referring the reader
to I for the details.

\subsection{Short review of the formalism}

The starting point in the second-order formalism is the following factorisation of the $x$-space Dirac propagator $S^{x'x}[A]$ in a Maxwell background (we suppress spin indices for brevity), 
\bear
S^{x'x}[A] &=&
\bigl[m + i\slash{D}'\bigr]
K^{x'x}[A]\, ,
\label{StoK}
\ear
where $\slash D^{\prime} = \gamma^{\mu}D^{\prime}_{\mu}$, $D_{\mu}^{\prime} = \partial_{\mu}^{\prime} + ieA_{\mu}(x') $ is the covariant derivative and we introduced the matrix elements\footnote{See appendix \ref{app-conv} for our conventions.}
\bear
K^{x'x}[A] &\equiv &
\langle x' \big| \bigl[m^2 + \jhat{\pi}^\mu \jhat{\pi}_\mu +{i\over 2}\, e \gamma^{\mu}\gamma^{\nu} F_{\mu\nu} ( \jhat{x})\bigr]^{-1} 
\big| x \rangle
\label{defK}
\ear
where $\jhat{\pi}_\mu = \jhat{p}_\mu +e A_\mu(\jhat{x})$.
For this ``kernel'' function, following \cite{fragit}, we then derived the following path integral representation:
\bear
K^{x'x}[A] &=&
\int_0^{\infty}
dT\,
\e^{-m^2T}
\e^{-\fourth \frac{(x-x')^2}{T}}
\int_{q(0)=0}^{q(T)=0}
Dq\,
{\rm e}^{-\int_0^T d\tau\bigl[
\kinq
+ie\,\dot q\cdot A (x_0+q)
+ie \frac{x'-x}{T}\cdot A (x_0+q)
\bigr]}
\nonumber\\
&& \times 2^{-\frac{D}{2}}
{\rm symb}^{-1}
\int_{\psi(0)+\psi(T)=0
\hspace{-30pt}}D\psi
\, \e^
{-\int_0^Td\tau\,
\bigl[\half\psi_{\mu}\dot\psi^{\mu}-ie(\psi+\eta)^{\mu}F_{\mu\nu}(x_0+q)(\psi+\eta)^{\nu}\bigl]
}\, .
\label{Kfin}
\ear\no
Here $x_0= x+ \frac{\tau}{T}(x'-x)$ is the straight-line path between the endpoints, $\eta^{\mu}$ is an external Grassmann Lorentz vector, and the inverse of the ``symbol map'' {\it symb} converts products of $\eta$'s  into fully antisymmetrised products of Dirac matrices,
\bear
{\rm symb} 
\bigl(\gamma^{[\alpha_1\alpha_2\cdots\alpha_n]}\bigr) \equiv 
(-i\sqrt{2})^n
\eta^{\alpha_1}\eta^{\alpha_2}\ldots\eta^{\alpha_n}\, .
\label{defsymb}
\ear
When working in four dimensions, this reduces to the three cases
\bear
{\rm symb}^{-1}(1) &=& \Eins \, ,\\
{\rm symb}^{-1}(\eta^{\alpha_1}\eta^{\alpha_2}) &=& - \frac{1}{4} \lbrack \gamma^{\alpha_1},\gamma^{\alpha_2}\rbrack \, ,\\
{\rm symb}^{-1}(\eta^{\alpha_1}\eta^{\alpha_2}\eta^{\alpha_3}\eta^{\alpha_4}) &=& 
 - \frac{i}{4}\varepsilon^{\alpha_1\alpha_2\alpha_3\alpha_4}\gamma_5\, .
\label{evalsymb}
\ear 
We then performed the usual projection onto an $N$-photon background of fixed momenta and polarisations, and used Gaussian integration on both the orbital
path integral, $\int Dq(\tau)$, and spin path integral, $\int D \psi(\tau)$,
to derive Bern-Kosower type master formulas for the $N$-photon kernel $K$ both in configuration and in momentum space. We also showed a convenient way to decompose these amplitudes into contributions with a fixed number of orbital and spin interactions, which we refer to as a spin-orbit decomposition. 

As in I, in our applications here we will focus entirely on momentum-space amplitudes. 
Fourier transforming the starting identity \eqref{StoK} to momentum space, and projecting it onto the $N$-photon sector, it turns into
\bear
S_N^{p'p} [k_1,\varepsilon_1;\ldots;k_N,\varepsilon_N]  &=& ({\slash p'}+m) K_N^{p'p}[k_1,\varepsilon_1;\ldots;k_N,\varepsilon_N] 
\nonumber\\ &&
- e\sum_{i=1}^N{\slash \varepsilon_i} K_{(N-1)}^{p'+k_i,p}[k_1,\varepsilon_1;\ldots;\hat k_i,\hat\varepsilon_i; 
\ldots; k_N,\varepsilon_N] \, .
\label{SN}
\ear
Here in the second term the `hat' on $\varepsilon_i$ and $k_i$ means omission. 
The terms involving $K_{(N-1)}$ are called ``subleading'' and arise because one of the $N$ photons could be taken from
the $A_{\mu}$ appearing in the covariant derivative in the factor $\bigl[m + i\slash{D}'\bigr]$ in \eqref{StoK}. An important advantage of the ``worldline representation'' is that it automatically takes care of the permutations over the $N$ photons,
and thus avoids the break-up of the amplitude into individual Feynman diagrams. This may not seem very relevant at the 
tree-level, but becomes an important issue when tree-level amplitudes are used for the construction of multiloop
amplitudes by sewing \cite{15,41,100}.     

\subsection{The on-shell case}
In I our focus was on the use of the off-shell amplitudes as a building block for loop amplitudes, which we
exemplified by a recalculation of the one-loop electron self-energy in an arbitrary gauge and dimension.
Now, our objective is to explore the simplifications which can be achieved in the on-shell case, both for the amplitudes themselves as well
as for the linear ($N=2$) and non-linear ($N> 2$) Compton scattering cross sections that can be constructed out of them. 
Thus our principal object of interest is the on-shell matrix element
corresponding to the dressed electron propagator with fixed spins $s,s'$.
In its construction we must remember that the second-order representation of the dressed electron propagator
still contains the external propagators, which must be removed before going on-shell.
Thus the matrix element has to be written as
\bear
{\cal M}_{N s^{\prime}s}^{p^{\prime} p}	
	=
	\bar{u}_{s'}(-p') (-{\slash p'}+m) S_N^{p'p} (\slash p +m) u_{s}(p)\,,
	\label{calMs}
\ear
where the spinors satisfy the on-shell relations
\bear
\bar{u}_{s'}(-p')(-{\slash p'}+m) = 0 = (\slash p +m) u_{s}(p) \, .
\label{onsh}
\ear
Thus these zeroes must be cancelled by corresponding poles to get a non-vanishing matrix element, which is the
fermionic version of the LSZ theorem. 
Looking at the decomposition \eqref{SN}, we can see how this works for the leading term: 
as explained in part I, $K_N^{p'p}$ is, in the second-order formalism,
built from untruncated Feynman diagrams involving only scalar propagators. Thus it contains a factor 
$\frac{1}{\left(p'^{2} + m^{2}\right)\left(p^2 + m^{2}\right)}$, which led us to the redefinition (I. 5.11),
\bear
	K_N^{p^{\prime}p} &=&  (-ie)^N \frac{\mathfrak{K}_N^{p^{\prime}p}}{\left(p'^{2} + m^{2}\right)\left(p^2 + m^{2}\right)}\, .
\label{defcalK}
\ear\no
Since $m^2+p^2= (\slash p +m)(-\slash p + m)$, the cancellation of the poles can then be made explicit: 

\bear
&&\bar{u}_{s'}(-p') (-{\slash p'}+m) (\slash p' +m)\frac{\mathfrak{K}_N^{p^{\prime}p}}{\left(p'^{2} + m^{2}\right)\left(p^2 + m^{2}\right)}
 (\slash p +m) u_{s}(p) 
\non
&& =
 \bar{u}_{s'}(-p') \frac{\mathfrak{K}_N^{p^{\prime}p}}{-\slash p + m} u_{s}(p) 
 =
  \bar{u}_{s'}(-p') \frac{\mathfrak{K}_N^{p^{\prime}p}}{2m} u_{s}(p) \, .
 \label{Knopole}
\ear
The same does not work for the subleading terms, $K_{(N-1)}^{p' + k_{i}, p}$, since in those the pole $\frac{1}{p'^2+m^2}$ has been shifted to 
$\frac{1}{(p'+k_i)^2+m^2}$. Thus they can be omitted in the calculation of on-shell matrix elements, and our
final formula for the matrix elements of the $N$-photon dressed electron propagator becomes
\bear
{\cal M}_{N s^{\prime}s}^{p^{\prime} p}	
	&=& \frac{(-ie)^N}{2m}
	 \bar{u}_{s'}(-p') \mathfrak{K}_N^{p^{\prime}p} u_{s}(p) \, .
	\label{calMsspinfin}
\ear
As is clear from \eqref{evalsymb}, in $D=4$ the Dirac matrix structure of $\mathfrak{K}_N^{p^{\prime}p}$ is (I 5.11),
\bear
\mathfrak{K}_N^{p^{\prime}p} &=& 
		 A_N \bone + B_{N\ab}\sigma^{\alpha\beta} -i C_N\gamma_{5} \, ,
\label{defABC}
\ear\no
with $\sigma^{\alpha\beta} = \frac{1}{2} [\gamma^\alpha , \gamma^\beta ]$. Moreover,  from the spin-orbit
decomposition it is clear that $A_N$ can be split as
\bear
A_N =  A_N^{\rm scal}+A_N^{\psi}\, ,
\label{splitA}
\ear
where $A_N^{\rm scal}$ is, up to the coupling constant factor, the truncated scalar propagator (see (I.1.2) for its path integral representation), 
\bear
(-ie)^N A_N^{\rm scal} =  \jhat{D}_N^{p'p}\, ,
\label{defANscal}
\ear
while terms in $A_N^{\psi}$ involve the spin interaction at least once. 

On-shell matrix elements in QED are guaranteed to be transversal in the photon polarisations, implying that it must
be possible to write the amplitude entirely in terms of photon field-strength tensors, defined by
\begin{equation}
	f_{\mu\nu} = k_{\mu}\varepsilon_{\nu} -\varepsilon_{\mu} k_{\nu}\,.
	\label{eqPhotonf}
\end{equation} 
for a photon of momentum $k$ with polarisation vector $\varepsilon$.
Although this fact is even stated in some
textbooks on QFT (see, e.g., \cite{peskinschroeder-book}), in the standard Feynman diagram approach it is by no means trivial 
to actually achieve this rewriting explicitly for an arbitrary number of photons (see, e.g., \cite{felihu}). 
Here we will show  both for the dressed scalar propagator $D_N^{p'p}$ as well as for the kernel $K_N^{p'p}$ of the
dressed electron propagator how this manifest transversality can be achieved at the integrand level by a simple integration-by-parts algorithm. 
We will also develop some useful formulas for the application of the standard spinor helicity technique to expressions where all
polarisation vectors are contained in field-strength tensors. 

A less obvious simplification concerning the coefficients $A_N,B_{N\mn},C_N $ is that there are useful relations between them
whenever both $p$ and $p'$ are on-shell. We will see that, for $D=4$ ($\widetilde{B}$ is the dual to $B$), 
\bear
A_N(m^2-p\cdot p') &=& 2p^\mu B_{N\mu\nu} p'^\nu \label{intro-idAB2}\\
C_N(m^2+p\cdot p')&=& 2p^\mu  \widetilde B_{N\mu\nu}p'^\nu\, , \label{intro-idCB2}
\ear
so that in principle the full information on the matrix element is already contained in $B_{N\mu\nu}$. 

Our goal of obtaining general expressions for the fully polarised matrix elements, valid for any $N$ without the explicit
knowledge of the coefficient functions $A_N,B_{N\mn},C_N$, made it necessary to dispose of explicit formulas for the Dirac
bilinears appearing when \eqref{defABC} is put into \eqref{calMsspinfin}. To our surprise, we could not find in the literature
explicit formulas for those bilinears valid for general $p$, $p'$ and arbitrary spin axes. We have therefore derived suitable formulas
ourselves, based on an algorithm proposed recently in \cite{olpozp} (see section \ref{sec:bilinears}). 

After summing over electron spins, which we are still able to do for arbitrary $N$ and photon helicity
assignments, further simplification is possible, leading to the following very compact representation for the spin-averaged
cross sections:
\bear
\big\langle \big\pipe{\cal M}_N^{p'p}\big\pipe^2\big\rangle &=& e^{2N} \Bigl\lbrack \abs{A_N}^2 + 2 B_{N}^\ab B_{N\ab}^{\ast}
 - \abs{C_{N}}^2
 \Bigr\rbrack .
 \label{optimal}
\ear
We then apply all this machinery to the $N=2$ case, performing a complete recalculation of the fully polarised 
matrix elements for Compton scattering, as well as the spin-averaged and the fully unpolarised cross sections, and
recover the known results with relatively little effort. 

The organisation of part 2 is as follows. Sections 2 - 6 are preparatory: in section \ref{R} we explain the on-shell IBP procedure leading to the
``R-representation'' of the integrands of the on-shell photon-dressed scalar and electron propagator. 
Section \ref{phopol} contains a collection of formulas involving the fixed-helicity field-strength tensors $f^{\pm\mn}$.
As a warm-up, in section \ref{examples} we apply the formalism to the scalar QED case,
obtaining the $N=2$ matrix elements and demonstrating the well-known vanishing of the ``all +'' amplitudes in the massless
limit. 
Section \ref{sec:bilinears} is devoted to the construction of Dirac bilinears, section \ref{sec:relations} to the study
of the coefficient functions $A_N,B_{N\mn},C_N$.
The central section of this paper is \ref{sec:matel}, which contains the construction of the fully polarised matrix elements
as well as the spin-averaged cross sections, for arbitrary photon number and helicity assignments. 
Section \ref{sec:allplus} deals with the ``all +'' case, which exhibits some further interesting simplifications
related to the well-known connections between helicity, self-duality and supersymmetry 
\cite{thooft,dadda,brolee,dufish1,dufish2,51}. 
In section \ref{sec:compton} we apply all these developments to the ordinary Compton scattering case. 
Our conclusions and directions of future work are offered in section \ref{sec:conc}. 

There are four appendices. In appendix \ref{app-conv} we summarise our conventions; contrary to I, where Euclidean conventions were used throughout, for the computation of on-shell
quantities it is, as usual, preferable to Wick rotate from Euclidean to Minkowski space. 
Apart from changing the metric from $(++++)$ to $(-+++)$, this will also induce a factor of $(-i)$ for each propagator and a factor of $i$ for each vertex. This amounts to a global factor of $(-i)$ for the dressed scalar or spinor propagators 
$D_N^{p'p}$ or $S_N^{p'p}$, but a factor $i$ for their truncated versions $\jhat D_N^{p'p}$ and $\jhat S_N^{p'p}$, defined by
\bear
\jhat D_{N}^{p'p} &\equiv &   \left(p^{\prime 2} + m^{2}\right)D_{N}^{p^{\prime}p}	 \left(p^{2} + m^{2}\right)\, , \label{defhatD}\\
\jhat S_N^{p'p} &\equiv& (-{\slash p'}+m) S_N^{p'p} (\slash p +m)\, . \label{defhatS}
\ear
Thus in Minkowski space we have $i{\cal M}_{{\rm scal},N}^{p'p} = i \jhat D_N^{p'p}$
and $i{\cal M}_{{\rm spin},N}^{p'p} = i \jhat S_N^{p'p}$. 

In appendix \ref{app-bispecial} we provide explicit formulas for the Dirac bilinears of section \ref{sec:bilinears} for the two 
most standard choices of electron polarisations, projection of the spin on the direction of motion or on the $z$-axis in the electron's rest frame. Appendix \ref{app-recursion} derives recursion formulas for the coefficient functions $A_N,B_{N\mn},C_N$ that provide an
alternative algorithm for their determination. Finally some simplifications peculiar
to the massless limit are discussed in appendix \ref{app-massless}. 
 
\section{Manifest transversality at the integrand level}
\label{R}

An essential feature of the worldline formalism is the option of using integration-by-parts (`IBP') \cite{strassler1, strassler2, 41, 26, 91} to improve on the integrand of parameter integrals before their actual calculation.
For the prototypical case, the one-loop $N$-photon amplitudes, one can hope to reduce significantly the
number of terms in the integrand, to achieve manifest gauge invariance at the integrand level,
eliminate spurious UV divergences (for the $N=4$ case), and realise the unification of the scalar and spinor QED cases
through the ``Bern-Kosower cycle replacement rule'' \cite{berkosPRL,berkosNPB}. 
Although similar manipulations could also be done using Feynman diagrams, they would usually become much more cumbersome
as well as harder to find since they do not involve individual Feynman diagrams but rather whole sets of them. 

For general $N$, two different IBP algorithms have already been developed:
``symmetric IBP'' \cite{26,41} which leads to the ``Q-representation,''
 well-suited to the application of the ``cycle replacement rule,'' and the algorithm of \cite{91}
resulting in the ``R-representation,'' which provides manifest gauge invariance (it is also possible to combine the two \cite{91}). 

Here in the open-line case, this becomes a matter of ``off-shell'' vs. ``on-shell,''
since the IBPs leading to the R-representation generate boundary terms that in the on-shell case can be omitted due to the LSZ theorem, but would be messy to deal with in the off-shell case; this is why in part I we focused on ``symmetric IBP,'' while for our
present on-shell purposes we prefer to work with the R-representation. Thus we will now derive this representation,
starting with the scalar line and then using the spin-orbit decomposition to extend it to the spinor line. 

\subsection{Manifestly transversal representation of the dressed scalar propagator}
\label{transD}

The basic idea of the R-representation has been explained already in section 2.3 of I. The electromagnetic potential is specialised to a sum of plane waves, $A^{\mu}(x) = \sum_{i=1}^{N}\veps_{i}^{\mu} \e^{i k_{i} \cdot x}$, and we project onto the contribution to the kernel that is multi-linear in the $\veps_{i}$. The photons inserted along the line are then represented by vertex operators,
\bear 
V_{\rm scal}[k,\varepsilon]=\int_0^Td\tau\, \veps\cdot\dot{x}(\tau)\,{\rm e}^{ik\cdot x (\tau)} \, .
\label{defvertop}
\ear
The simplest way of achieving manifest transversality of photonic amplitudes is to
rewrite each vertex operator from the beginning in terms of the photon field-strength tensor.
In scalar QED one can do this by adding to the vertex operator a total-derivative term supplying the ``missing half'' of that tensor:
\bear
V_{\textrm{scal}}[k, \veps; r] := V_{\rm scal}[k,\veps] 
+ i  \frac{\veps\cdot r}{k \cdot r} \int_0^Td\tau \frac{d}{d\tau} \e^{ik\cdot x(\tau)}
=
\int_0^Td\tau \frac{r\cdot f\cdot {\dot x}}{r\cdot k} \e^{ik\cdot x(\tau)}\, .
\label{defVertopmod}
\ear
Here $r$ is a ``reference vector'' that is arbitrary except for the constraint $k\cdot r \ne 0$.
In the open-line case, adding this term causes non-vanishing boundary terms, but these do not 
have both of the poles necessary for contributing to the on-shell matrix element ${\cal M}_{{\rm scal},N}^{p'p}$, defined
by the on-shell limit of
$\left(p^{\prime 2} + m^{2}\right)D_{N}^{p^{\prime}p}	\left(p^{2} + m^{2}\right)$. 
Thus, as far as the calculation of ${\cal M}_{{\rm scal},N}^{p'p}$ is concerned, 
the $N$-photon dressed scalar propagator 
$D_{N}^{p^{\prime}p}$ could as well be used with the replacement
\bear
\varepsilon_i \rightarrow \frac{r_i\cdot f_i}{r_i\cdot k_i}, \quad i=1,\ldots ,N
\ear
throughout. Thus, in particular, the master formula (I. 2.25) (originally due to \cite{dashsu} and \cite{ahmbassch-16}),
\bear 
\hspace{-1em}D_N^{p'p}(k_1,\veps_1;\cdots;k_N,\veps_N) &=& (-ie)^N
\int_0^\infty dT\, {\rm e}^{-m^2T}
\nonumber\\
\hspace{-1em} &\times&
\int_0^T\prod_{i=1}^{N} d\t i\,
{\rm e}^{\sum_{i,j=0}^{N+1}\big[\frac{1}{2}\vert \tau_i-\tau_j\vert
k_i\cdot k_j -i {\rm sgn}(\tau_i-\tau_j)\varepsilon_i\cdot
k_j+\delta (\tau_i-\tau_j)\epseps ij \big]}
\multilin \, ,
\nonumber\\ \label{master-dss} \ear
$
(k_0 = p'\, , 
k_{N+1} = p\,,
\tau_0 =  T \, , 
 \tau_{N+1}  =   
 \varepsilon_0 = \varepsilon_{N+1} = 0 )
 $
can then as well be written purely in terms of the field strength tensors of the $N$ photons in the form
\begin{align}
\hspace{-2em}D_N^{p'p}(k_1,\veps_1,r_1;\cdots;k_N,\veps_N,r_N)  &= (-ie)^N
\int_0^\infty dT\, {\rm e}^{-m^2T}
\nonumber\\ 
\hspace{-2em}&\times
\int_0^T\prod_{i=1}^{N} d\t i\,
{\rm e}^{\sum_{i,j=0}^{N+1}\big[
\vert \tau_i-\tau_j\vert \half  k_i\cdot k_j
-i {\rm sgn}(\tau_i-\tau_j){r_i\cdot f_i\cdot k_j\over r_i\cdot k_i}
- \delta (\tau_i-\tau_j) 
{r_i\cdot f_i\cdot f_j\cdot r_j\over r_i\cdot k_i \, r_j\cdot k_j}
\big]}
\Bigg\vert_{f_1f_2\ldots f_N}\nonumber \\
\label{linemastercov} 
\end{align}
even though the right-hand sides of \eqref{master-dss} and \eqref{linemastercov} 
are not actually equal. 

The result of expanding out the exponential factor in (\ref{linemastercov}) will be named
$(-i)^N \bar R_N\, {\rm e}^{(\cdot)}$, 
where ${\rm e}^{(\cdot)}$ was already introduced in part I,
\bear 
{\rm e}^{(\cdot)}&\equiv& {\rm e}^{\half\sum_{i,j=0}^{N+1}  \vert \tau_i - \tau_i \vert  k_i\cdot k_j} \nonumber\\
&=& {\rm e}^{ -Tp'^2 + \frac{1}{2}\sum_{i,j=1}^N  \vert \tau_i - \tau_i \vert k_i\cdot k_j + (p-p') \cdot \sum_{i=1}^N k_i \tau_i } \, .
\label{expred}
\ear 
The on-shell master formula (\ref{linemastercov}) can then be rewritten as
\bear
D_N^{p'p}(k_1,\veps_1,r_1;\cdots;k_N,\veps_N,r_N) &=& (-e)^N \int_0^\infty dT\, {\rm e}^{-m^2T}
\int_0^T\prod_{i=1}^{N} d\t i\ \bar R_N\, {\rm e}^{(\cdot)}. \label{Master_es_onshell}
\ear
For use below let us also write down the first two polynomials $\bar R_N$:
\begin{align}
\begin{split}
\hspace{-1em}\bar R_1 &= \sum_{i=0}^2
{\rm sgn}(\tau_1-\tau_i)\frac{r_1\cdot f_1\cdot k_i}{r_1\cdot k_1}
=
\frac{r_1\cdot f_1\cdot (p-p')}{r_1\cdot k_1}
\, ,
\\
\hspace{-1em}\bar R_2 &= \sum_{i,j=0}^3
{\rm sgn}(\tau_1-\tau_i)
\frac{r_1\cdot f_1\cdot k_i}{r_1\cdot k_1}
{\rm sgn}(\tau_2-\tau_j)
\frac{r_2\cdot f_2\cdot k_j}{r_2\cdot k_2}
+ 2\delta(\tau_1-\tau_2)
{r_1\cdot f_1\cdot f_2\cdot r_2\over r_1\cdot k_1 \, r_2\cdot k_2}\,.
\end{split}
\label{barR1R2}
\end{align}
To obtain additional simplifications, we observe that the most complicated term in $\bar R_N$ is always given by the product
\bear
\prod\limits_{j=1}^N \Bigl(\sum_{i_{j}=0}^{N +1}
{\rm sgn}(\tau_j-\tau_{i_{j}})
\frac{r_j\cdot f_j\cdot k_{i_{j}}}{r_j\cdot k_j}
\Bigr)
 \label{gdot_t}
\ear
which, upon choosing\footnote{Excluding the case when $N=1$, since here $p' \cdot k_1 = 0$ on-shell.} $r_j = p'$ for
$j=1,\ldots,N$, turns into

\bear
\prod\limits_{j=1}^N \frac{ p' \cdot f_j\cdot \bigl( p +\sum_{i_{j}=1}^N {\rm sgn}(\tau_j-\tau_{i_{j}})
 k_{i_{j}} \bigr) }{p' \cdot k_j}\, . \label{gdot_t2}
\ear
For every ordering of the variables $\tau_1,\ldots,\tau_N$, there is a $j$ such that the $j$th term in the product is given by 
\begin{align}
\hspace{-0.5em}p' \cdot f_j\cdot \Bigl( p +\sum_{i_{j}=1}^N {\rm sgn}(\tau_j-\tau_{i_{j}}) k_{i_{j}} \Bigr) =
 p' \cdot f_j \cdot \Bigl(p + \sum_{i_{j}=1, i_{j}\neq j}^N k_{i_{j}} \Bigr) 
= - p' \cdot f_j \cdot ( p' + k_j ) = 0\, .
\end{align}
Thus the term in question can be made to vanish by this choice of reference momenta, leaving only those contributions in $\bar R_N$ that involve at least one of the $\delta(\tau_i-\tau_j)$ (these $\delta$-functions generate the seagull vertex of scalar QED). 

However, it is important to stress that the reference momenta $r_i$ cannot be chosen term-by-term, but rather have to be fixed
globally. Their role is, in fact, very similar to the reference momenta of the spinor helicity formalism
to be used below, except that they are not restricted to be null vectors. 
 
\subsection{Manifestly transversal representation of the N-photon kernel}
\label{transK}
Proceeding to the spinor-line case, in section 6 of part I we had seen that the $N$-photon kernel $K_N^{p'p}$ can be decomposed naturally as
\bear
K_N &=& \sum_{S=0}^N K_{NS}\, ,	
\label{decompso}
\ear
since in the representation \eqref{Kfin} after specialising to an $N$-photon background a certain number $S$ of photons have to be taken out of the spin path integral $\int D\psi$, and the remaining $N-S$ ones out of the coordinate (orbital) path integral $\int Dq$. 
The former ones are by convention associated to the variables $\tau_1,\ldots, \tau_S$, and 
since the spin path integral couples to the background field only through the field-strength tensor $F_\mn$, manifest transversality in those indices is automatic, as is also borne out by our explicit formula for these terms, written in terms of functions $W_{\eta}$, (6.8) of part I (some of these are given below). Thus only the transversality of the orbital part has to be taken care of, and here it is obvious that the only change necessary in the final formula for $K_{NS}$, (I. 6.10), is to replace the 
orbital prefactor polynomials $\bar P_{NS}^{\lbrace i_1i_2\ldots i_S\rbrace}$ by the corresponding objects of the R representation,
defined by (compare \eqref{linemastercov} and (I.6.11))
\begin{align}
{\rm e}^{\sum_{i,j=0}^{N+1}\big[
-i {\rm sgn}(\tau_i-\tau_j){r_i\cdot f_i\cdot k_j\over r_i\cdot k_i}
- \delta (\tau_i-\tau_j) 
{r_i\cdot f_i\cdot f_j\cdot r_j\over r_i\cdot k_i \, r_j\cdot k_j}
\big]}
\big \vert_{f_{i_1}= \cdots = f_{i_S} =0}\big \vert_{f_{i_{S+1}}\cdots f{i_N}}
&\equiv &
(-i)^{N-S} \bar R_{NS}^{\lbrace i_1i_2\ldots i_S\rbrace } \,.
\label{defRNS}
\end{align}
Thus we arrive at the following manifestly transversal version of the spin-orbit decomposition (compare with (I 6.10)):
\bear
K_{NS} &=& \sum_{\lbrace i_1i_2\ldots i_S\rbrace} K_{NS}^{\lbrace i_1i_2\ldots i_S\rbrace}\, , \nonumber\\
K_{NS}^{\lbrace i_1i_2\ldots i_S\rbrace }  &=& (-e)^N \symbi\int_0^\infty dT\,{\rm e}^{-m^2T}\prod_{i=1}^N\int_0^Td\tau_i\,
W_{\eta} (k_{i_1},\veps_{i_1};\ldots;k_{i_S},\veps_{i_S}) 
\bar R_{NS}^{\lbrace i_1i_2\ldots i_S\rbrace} 
{\rm e}^{(\cdot)}\, ,
\nonumber\\
\label{KNSexpl}
\ear
where the sum 
runs over all choices of $S$ out of the $N$ variables, and ${\rm e}^{(\cdot)}$ was given in \eqref{expred}.

For example, the spin-orbit decomposition of the two-photon kernel will now, instead of (I.6.13), take the form
\begin{align}
K_2^{p'p} &= e^2  {\rm symb}^{-1} \int_0^\infty dT\, \e^{-(m^2+p'^2)T}\int_0^T d\tau_1 d\tau_2\,  \e^{k_1\cdot k_2 |\tau_1-\tau_2|-(p'-p)\cdot (\tau_1k_1+ \tau_2 k_2)}
\nonumber\\
&\times \Bigl\{ \bar R_2 +    W_{\eta}(k_1,\veps_1)\bar R_{21}^{\lbrace 1\rbrace}+ W_{\eta}(k_2,\veps_2)\bar R_{21}^{\lbrace 2\rbrace}
+ W_{\eta} (k_1,\veps_1;k_2,\veps_2)
\Bigr\}~.
\label{K2decomp}
\end{align}
Here $\bar R_2$ was given in \eqref{barR1R2}, while
\begin{align}
\begin{split}
\bar R_{21}^{\lbrace1\rbrace} &= \frac{r_2\cdot f_2 \cdot [p-p'+{\rm sgn}(\tau_2-\tau_1)k_1]}{r_2\cdot k_2}
\, , \\
\bar R_{21}^{\lbrace2\rbrace} &=  \frac{r_1\cdot f_1\cdot [p-p'+{\rm sgn}(\tau_1-\tau_2)k_2]}{r_1\cdot k_1}
\, ,
\end{split}
\label{barR21}
\end{align}
and
\begin{align}
\begin{split}
W_{\eta} (k_i,\veps_i) &=  \eta f_i \eta \, ,\\
W_{\eta} (k_1,\veps_1;k_2,\veps_2) &=  \frac{1}{2} \tr (f_1f_2)+
 2G_{F12}\eta f_1f_2\eta + \eta f_1 \eta\, \eta f_2 \eta\, .
\end{split}
\end{align}
We shall use these explicit results below when we recover the standard Compton cross section. 
\section{Photon polarisations and spinor helicity}
\label{phopol}

As is usual in the high-energy physics context, we will fix the on-shell photon polarisations using a basis of 
helicity eigenstates, and employ the spinor helicity formalism for the construction of these states. Following the conventions of 
\cite{srednicki-book,elvhua-book}, they are given by 

\bear
\varepsilon^{+\mu}(k;q)= - \frac{\langle q\pipe \gamma^{\mu}\pipe k\rbrack}
{\sqrt{2}\langle q k \rangle}\, , \qquad \varepsilon^{-\mu}(k;q)= - \frac{\lbrack q\pipe \gamma^{\mu}\pipe k\rangle}
{\sqrt{2}\lbrack q k \rbrack} \, ,
\label{defpolpm}
\ear
where $k$ denotes the photon momentum and $q$ the reference vector. 
Since in our representation all polarisation vectors are already absorbed into field-strength tensors, let us also collect here a number of formulas involving field strength tensors of on-shell photons with fixed circular polarisations,
\bear
f^{\pm \mn} &\equiv& k^{\mu}\varepsilon^{\pm \nu} - \varepsilon^{\pm \mu} k^{\nu}
\label{deffpm}
\ear
(some but not all of these formulas were already given in \cite{56}). 

First, since the $f^{\pm\mn}$ are transversal they should be independent of the reference momentum $q$. And indeed, 
one can show that they can be written purely in terms of $k$ as follows,
\begin{equation}  \begin{aligned}
f^{+\mn} &= 
-\frac{1}{4\sqrt{2}} [k|[\gamma^{\mu},\gamma^{\nu}]|k] \, ,\qquad
f^{-\mn} &= -\frac{1}{4\sqrt{2}} \langle k|[\gamma^{\mu},\gamma^{\nu}]|k\rangle \, .
   \end{aligned} \label{fpmapp}\end{equation} 

The following matrix identities can then easily be established:

\begin{itemize}

\item
{\it Hermitian conjugation:}
\bear
f^{+\dag} &=& f^{-} \, .
\ear

\item
{\it (Anti-) self duality:} 
\bear
\widetilde f^{\pm } &=& \pm i f^{\pm } 
\label{idfdual}
\ear
where $\widetilde f^{\alpha\beta} \equiv \half\varepsilon^{\alpha\beta\gamma\delta}f_{\gamma\delta}$
is the dual field-strength tensor. 

\item
{\it Products:}
\bear
\bigl(f_1^+f_2^+\bigr)^{\mu\nu} &=&
\frac{1}{4} [12][1|\gamma^\mu\gamma^\nu |2] 
\, ,
\\
\bigl(f_1^-f_2^-\bigr)^{\mu\nu} &=&
\frac{1}{4} \langle 1 2 \rangle \langle 1 |\gamma^\mu\gamma^\nu |2\rangle
\, , \\
\bigl(f_1^+f_2^-\bigr)^{\mu\nu} 
=
\bigl(f_2^-f_1^+\bigr)^{\mu\nu} 
&=&
\fourth\,
[1\vert\gamma^{\mu}\vert 2\rangle
[1\vert\gamma^{\nu}\vert 2\rangle 
\, .
\label{F1+F2-}
\ear
Here and in the following we will often follow the usual notation of replacing
a $k_i$ by $i$ inside spinorial expressions. 

\item
{\it Anticommutators:}
\bear
\lbrace f_1^+,f_2^+ \rbrace^{\mn} &=& -\half [12]^2\eta^{\mn}\, ,\label{anticomm++} \\
\lbrace f_1^-,f_2^- \rbrace^{\mn} &=& -\half \langle 12\rangle^2\eta^{\mn} 
\, .
\ear

\item
{\it Factorisation of traces:}
\bear
\tr (f^+_{i_1}\cdots f^+_{i_M}f^-_{j_1}\cdots f^-_{j_N}) 
= \fourth\tr (f^+_{i_1}\cdots f^+_{i_M})\tr (f^-_{j_1}\cdots f^-_{j_N})
\, .
\ear

\item
{\it Same-helicity traces:} 
\bear
{\rm tr} (f^+_{i_1}\cdots f^+_{i_N}) &=&  \dfrac{(-1)^{N}}  {\sqrt{2^{N-2}}}   \; [i_1 i_2] \dots [i_{N-1} i_N] [i_N i_1] \, ,\label{plustraces}\\
{\rm tr} (f^-_{i_1}\cdots f^-_{i_N}) &=&  \dfrac{(-1)^{N}}{\sqrt{2^{N-2}}} \; \langle i_1 i_2 \rangle \dots \langle i_{N-1} i_N \rangle \langle i_N i_1 \rangle   \, .\label{minustraces}
\ear

\end{itemize}

Finally, let us also write down spinor helicity expressions for $\bar R_1$ and $\bar R_2$ for the case where the $r_i$ are 
null vectors:

\bear
\frac{r\cdot f^+\cdot k_i}{r\cdot k} = \frac{1}{\sqrt{2}} \frac{\langle r k_i\rangle \lbrack k_i  k \rbrack}{\langle r k\rangle} \,, \qquad 
\frac{r\cdot f^-\cdot k_i}{r\cdot k} =
\frac{1}{\sqrt{2}} \frac{\lbrack r k_i\rbrack \langle k_i  k \rangle}{\lbrack r k\rbrack} 
\, ,
 \ear

\bear
\frac{r_1\cdot f_1^+\cdot f_2^+\cdot r_2}{r_1\cdot k_1 r_2\cdot k_2} 
= - \frac{\langle r_1 r_2\rangle \lbrack k_1 k_2 \rbrack}{\langle r_1 k_1 \rangle \langle r_2 k_2 \rangle }
\, ,
 \\
\frac{r_1\cdot f_1^-\cdot f_2^-\cdot r_2}{r_1\cdot k_1 r_2\cdot k_2} 
= - \frac{\lbrack r_1 r_2\rbrack \langle k_1 k_2 \rangle}{\lbrack r_1 k_1 \rbrack \lbrack r_2 k_2 \rbrack } 
\, ,
\ear

\bear
\frac{r_1\cdot f_1^+\cdot f_2^-\cdot r_2}{r_1\cdot k_1 r_2\cdot k_2} 
=
\frac{\langle r_1k_2\rangle \lbrack k_1r_2\rbrack}{\langle r_1k_1\rangle \lbrack k_2 r_2 \rbrack}
\, ,
\label{rffrpm}\\
\frac{r_1\cdot f_1^-\cdot f_2^+\cdot r_2}{r_1\cdot k_1 r_2\cdot k_2} 
=
\frac{\lbrack r_1k_2\rbrack \langle k_1r_2\rangle}{\lbrack r_1k_1\rbrack \langle k_2 r_2 \rangle} 
\, .
\ear

\section{Some scalar-line calculations}
\label{examples}

As a warm-up, let us illustrate the use the formalism by rederiving some known results in scalar QED. 

\subsection{Vanishing of the ``all +'' amplitudes for the massless scalar line}
\label{secAll+}

For starters, let us rederive the fact that the scalar propagator dressed with any number of equal-helicity photons 
gives a vanishing amplitude in the limit of vanishing scalar mass (see, amongst others, \cite{Bernicot}).

This fact we can see most directly from the scalar master formula in its ``first alternative version,'' equation (I.2.23):
\begin{align}
\hspace{-1.5em}&D_N^{p'p}(k_1,\veps_1;\cdots;k_N,\veps_N) = (-ie)^N
\int_0^\infty dT\, {\rm e}^{-T(m^2+p'^2)}\int_0^T\prod_{i=1}^{N} d\tau_i\nonumber\\
&\times {\rm e}^{\sum_{i=1}^N(-2k_i\cdot p'\tau_i+2i\veps_i\cdot p')+\sum_{i,j=1}^N\big[\bigl(\frac{\vert \tau_i-\tau_j\vert}{2}-\frac{\tau_i+\tau_j}{2}\bigr)k_i\cdot k_j-i({\rm sgn}(\tau_i-\tau_j)-1)\veps_i\cdot k_j+\delta(\tau_i
-\tau_j)\veps_i\cdot\veps_j\big]}
\Big\vert_{\veps_1\veps_2\cdots \veps_N}\,. 
\label{scalarqedmasteropen}
\end{align}
If the scalar is massless, $p'$ is a null vector on-shell, so we can take it as the reference vector for all the photons.
This removes the terms proportional to $\varepsilon_i\cdot p'$ in the exponent. 
Then if all the photons have the same helicity, the identity 
\bear
\varepsilon^\pm(k_i;q)\cdot \varepsilon^\pm(k_j;q) = 0
\ear
removes all the terms in the exponent involving $\veps_i\cdot\veps_j$. 
The only terms left are of the form
\bear
\prod_{i=1}^N 
\Bigl({\rm sgn}(\tau_i-\tau_{j_i})-1\Bigr)\veps_i\cdot k_{j_i}.
\ear
Now we apply a similar argument to the one given below (\ref{barR1R2}): for any given ordering of the $\tau_i$ there will be one $\tau_{i_0}$ with the largest value, so that 
${\rm sgn} (\tau_{i_0} - \tau_{j_{i_0}}) = 1$, which makes the whole term vanish (taking also into account that on-shell we cannot have 
$j_{i_0} = i_0$ since $\veps_{i_0} \cdot k_{i_0} =0$). This completes the proof of the vanishing of this amplitude.

\subsection{N=2 amplitudes for the massive scalar line}
\label{subsec:massivescalar}

Next, let us have a look at the massive scalar line. In this case the vanishing theorem does not hold, so let us calculate
the two independent matrix elements for the $N=2$ case, ${\cal M}_{\rm{scal} \, 2}^{p'p++}$ and ${\cal M}_{\rm{scal} \, 2}^{p'p+-}$.

For the $++$ case, we use the on-shell master formula \eqref{Master_es_onshell} together with \eqref{barR1R2} 
and the observation that setting $r_1=r_2=p'$ removes the first term in $\bar R_2$, leading to 

\bear
D_2^{p'p}(k_1,\veps_1^+,p';k_2,\veps_2^+,p') &=& 2e^2
\frac{p'\cdot f_1^+\cdot f_2^+\cdot p'}{p'\cdot k_1 \, p'\cdot k_2}
 \int_0^\infty dT\, {\rm e}^{-(m^2+p'^2)T}
\int_0^Td\tau_1\int_0^Td\tau_2 \, \delta(\tau_1-\tau_2)
\nonumber\\
&& \times {\rm e}^{ \vert \tau_1 - \tau_2 \vert k_1\cdot k_2 + (p-p') \cdot ( k_1 \tau_1 + k_2 \tau_2 ) }\, .
\ear
The integrals simply give the propagators associated to the external legs,
\begin{equation}
\hspace{-1.5em}\int_0^{\infty}dT\,\e^{-(m^2+p'^2)T}\int_0^Td\tau_1\int_0^Td\tau_2 \, \delta(\tau_1-\tau_2)\,\e^{k_1\cdot k_2 \abs{\tau_1-\tau_2} - (p'-p)\cdot (\tau_1k_1+\tau_2 k_2)}
=  \frac{1}{(m^2+p^2)(m^2+p'^2)},, 
 \label{int2}
\end{equation}
since there is no internal propagator when the two photons meet at the same point along the line. Therefore
\bear
{\cal M}_{\rm{scal} \, 2}^{p'p++} = \jhat D_2^{p'p++} = (m^2+p^2)(m^2+p'^2) D_2^{p'p++} 
= 2e^2 \frac{p'\cdot f_1^+\cdot f_2^+\cdot p'}{p'\cdot k_1 \, p'\cdot k_2}\, .
\ear
Now we can use the identity \eqref{anticomm++}: 
\bear
p' \cdot f_1^{+} \cdot f_2^{+} \cdot p' &=& \frac{1}{2} \; p' \cdot \{ f_1^{+},f_2^{+} \} \cdot p' 
= - \frac{1}{4} \; [12]^2 \; p'^2 = \frac{m^2}{4} \; [12]^2\, ,
\ear
so that the final result becomes
\bear
{\cal M}_{\rm{scal} \, 2}^{p'p++} = 2 e^2 m^2 \frac{[12]^2}{\lambda_1\lambda_2}\, ,
\label{Mscal++fin}
\ear
in which we have further introduced the notation
\bear
\lambda_{i} &\equiv& 2p'\cdot k_{i} \, \qquad (i = 1, 2)\, .
\label{deflambda}
\ear
Turning now to the $+-$ component, we refine our choice of reference momenta according to
\bear
r_{i} &\equiv & p' + \frac{m^2}{\lambda_{i}} k_{i} \,  \quad (i = 1, 2)\, .
\label{defrpm}
\ear
These reference momenta are equally good as $p'$ for removing the first term in $\bar R_2$ as in (\ref{gdot_t2}), but have the advantage of being null vectors. Thus we can now apply \eqref{rffrpm} to write, in the second term,

\bear
\frac{r_1\cdot f_1^+\cdot f_2^-\cdot r_2}{r_1\cdot k_1 r_2\cdot k_2} 
=
\frac{\langle r_1k_2\rangle \lbrack k_1r_2\rbrack}{\langle r_1k_1\rangle \lbrack k_2 r_2 \rbrack}\, ,
\ear
leading to the compact form for the amplitude
\bear
{\cal M}_{\rm{scal} \, 2}^{p'p+-} = 2 e^2 \frac{\langle r_1k_2\rangle \lbrack k_1r_2\rbrack}{\langle r_1k_1\rangle \lbrack k_2 r_2 \rbrack}\, .
\label{Mscal+-fin}
\ear
This should be independent of the choice of reference momentum, and indeed it is easy to see that it can be rewritten as
\bear
{\cal M}_{\rm{scal} \, 2}^{p'p+-} = 2 e^2 
\frac
{\lbrack 1\vert \slash p' \vert 2 \rangle^2}
{\lambda_1\lambda_2} \;.
\label{Mscal+-finfin}
\ear

\section{Electron polarisations and Dirac bilinears}
\label{sec:bilinears}

In this section we develop some general formulae for Dirac bilinears that will be central in arriving at universal formulae for fully polarised amplitudes without the need to have explicit knowledge of the coefficients $A_{N}$, $B_{N\mu\nu}$ and $C_{N}$ or of the number or helicity of the photons participating in the scattering process. The on-shell electron polarisations can be fixed by imposing the following conditions \cite{srednicki-book},
\bear
\slash r \gamma_5 u_s(p) = s u_s(p), \qquad \slash z \gamma_5 u_{s'}(-p') = s' u_{s'}(-p'), \label{eq_u_pol}
\ear
with spin labels $s,s'=\pm$, and where the vectors $r^\mu$ and $z^\mu$, which define the directions for measuring the particle spins, satisfy $r^2=z^2=1$ and $r \cdot p=0 = z \cdot p'$. Using the approach outlined in \cite{olpozp}, we introduce a set of $4$-vectors with which we can construct quantities that span the space of the Dirac bilinears $\bar{u}_{s'}(-p') u_s(p)$, $\bar{u}_{s'}(-p') \sigma^{\mu\nu} u_s(p)$ and $\bar{u}_{s'}(-p') \gamma_5 u_s(p)$ appearing in \eqref{calMsspinfin} for the polarised amplitudes. 

To achieve this we introduce the null vector $d^\mu$ and its conjugate given with respect to the physical outgoing
momentum $-p^{\prime}$, as
\begin{align}
\begin{split}
d^\mu &= \dfrac{1}{4m} \overline{u}_+(-p') \gamma^\mu \gamma_5 u_-(-p')\, ,\\
d^{\ast\mu} &= \dfrac{1}{4m} \overline{u}_-(-p') \gamma^\mu \gamma_5 u_+(-p')\,.  
\end{split}
\label{Def_vectord}
\end{align}
These vectors are orthogonal to $-p'$ and $z$ and the set $\{-p', z, d, d^{*}\}$ forms a basis in the space of four-vectors. We further define the following four scalars,


\begin{align}
\begin{split}
\alpha_1 &= \dfrac{1}{4m^2} \left[ -p \cdot z \, r \cdot p' + (1+r \cdot z)(m^2+ p \cdot p' ) \right],
\\
\alpha_2 &= \dfrac{1}{2m^2} \left[ -p \cdot d^\ast \, r \cdot p' + r \cdot d^\ast (m^2+p\cdot p' ) \right], 
 \\
\alpha_3 &= \dfrac{1}{2m} \left[ p \cdot d^\ast ( 1 + r \cdot z ) - p \cdot z \, r \cdot d^\ast \right], 
\\
\alpha_4 &= -\dfrac{1}{4m} \left[ p \cdot z + r \cdot p' - 2 p \cdot d^\ast r \cdot d + 2 p \cdot d r \cdot d^\ast \right] \, ,
\end{split}
\label{DiracScalars}
\end{align}
and the normalisation factor
\bear
\mathcal{N} =  \dfrac{2m}{ \sqrt{\alpha_1 }}. \label{NormFactor}
\ear

One can now decompose the spinor $u_{s}(p)$ as a linear combination of the $u_{s'}(-p^{\prime})$ and their Dirac adjoints whose coefficients can be written in terms of the $\alpha_{i}$. The Dirac bilinears $\bar{u}_{s'}(-p') u_s(p)$ (scalar), and $\bar{u}_{s'}(-p') \gamma_5 u_s(p)$ (pseudoscalar) can then be expressed  in terms of these same scalars as follows,
\begin{align}
\overline{u}_+(-p') u_+(p) &= \mathcal{N} \alpha_1 , \quad & 
\overline{u}_+(-p') u_-(p) &= - \mathcal{N} \alpha_2^\ast , \nonumber \\
\overline{u}_-(-p') u_+(p) &= \mathcal{N} \alpha_2 , \quad & 
\overline{u}_-(-p') u_-(p)  &= \mathcal{N} \alpha_1,\nonumber \\
\overline{u}_+(-p') \gamma_5 u_+(p) &= - \mathcal{N} \alpha_4  , \quad &
\overline{u}_+(-p') \gamma_5 u_-(p) &= \mathcal{N} \alpha_3^\ast , \nonumber \\
\overline{u}_-(-p') \gamma_5 u_+(p) &= \mathcal{N} \alpha_3 , \quad &
\overline{u}_-(-p') \gamma_5 u_-(p) &= \mathcal{N} \alpha_4^\ast. \label{DiracBilinearAC}
\end{align}

Similarly, we can obtain the tensor bilinears decomposed in terms of antisymmetric combinations of the basis constructed above as
\bear
\overline{u}_{s'}(-p') \sigma_{\mu\nu} u_s(p) &=& 
\beta_{-p'z}^{s's} ( -p'_\mu z_\nu + z_\mu p'_\nu )
+ \beta_{-p'd}^{s's} ( -p'_\mu d_\nu + d_\mu p'_\nu ) \nonumber \\
&+& \beta_{-p'd^\ast}^{s's} ( -p'_\mu d^\ast_\nu + d^\ast_\mu p'_\nu)
+ \beta_{zd}^{s's} ( z_\mu d_\nu - d_\mu z_\nu ) \nonumber \\
&+& \beta_{zd^\ast}^{s's} ( z_\mu d^\ast_\nu - d^\ast_\mu z_\nu ) 
+ \beta_{dd^\ast}^{s's} ( d_\mu d^\ast_\nu - d^\ast_\mu d_\nu ), \label{DiracBilinearB}
\ear
where
\begin{align}
\beta_{-p'z}^{++} &= - \dfrac{\mathcal{N}}{m} \alpha_4, &
\beta_{-p'd}^{++} &= \dfrac{2 \mathcal{N}}{m} \alpha_3 , &
\beta_{-p'd^\ast}^{++} &= 0, \nonumber \\ 
\beta_{zd}^{++} &= - 2 \mathcal{N} \alpha_2, &
\beta_{zd^\ast}^{++} &= 0 , &
\beta_{dd^\ast}^{++} &= -2 \mathcal{N} \alpha_1, \nonumber \\ ~ \nonumber \\
\beta_{-p'z}^{+-} &= \dfrac{\mathcal{N}}{m} \alpha_3^\ast , &
\beta_{-p'd}^{+-} &=  \dfrac{2 \mathcal{N}}{m} \alpha_4^\ast, &
\beta_{-p'd^\ast}^{+-} &= 0 , \nonumber \\
\beta_{zd}^{+-} &= -2 \mathcal{N} \alpha_1, & 
\beta_{zd^\ast}^{+-} &= 0 , & 
\beta_{dd^\ast}^{+-} &= 2 \mathcal{N} \alpha_2^\ast , \nonumber \\ ~ \nonumber \\
\beta_{-p'z}^{-+} &= -\dfrac{\mathcal{N}}{m} \alpha_3 , &
\beta_{-p'd}^{-+} &= 0 , &
\beta_{-p'd^\ast}^{-+} &= - \dfrac{2 \mathcal{N}}{m} \alpha_4 , \nonumber \\
\beta_{zd}^{-+} &= 0 , &
\beta_{zd^\ast}^{-+} &= 2 \mathcal{N} \alpha_1 , &
\beta_{dd^\ast}^{-+} &= 2 \mathcal{N} \alpha_2 , \nonumber \\ ~ \nonumber \\
\beta_{-p'z}^{--} &= - \dfrac{\mathcal{N}}{m} \alpha_4^\ast , &
\beta_{-p'd}^{--} &= 0 , & 
\beta_{-p'd^\ast}^{--} &=  \dfrac{2 \mathcal{N}}{m} \alpha_3^\ast , \nonumber \\
\beta_{zd}^{--} &= 0 , &
\beta_{zd^\ast}^{--} &= -2 \mathcal{N} \alpha_2^\ast  , &
\beta_{dd^\ast}^{--} &= 2 \mathcal{N} \alpha_1.\label{DiracBeta}
\end{align}
In appendix \ref{app-bispecial} we compute the coefficients $\alpha_1, \ldots, \alpha_4$ explicitly for the two most common
choices of the spin axes (i) along the $z$-axis in the rest frame of the electron (ii) along the direction of motion. 

\section{Relations between the functions $A_N,B_{N\mn} ,C_N$}
\label{sec:relations}

Evaluating the dressed electron propagator $S_N$ between on-shell spinors allows us to exhibit hidden relations between the 
coefficient functions $A_N,B_{N\mn} ,C_N$ that hold whenever both electrons are on-shell. 
This works in the following way. 
Rather than factorising $S_N$ from the left, as we did in \eqref{SN}, we could as well have used the ``reversed'' identity (see (I.7.2))
\bear
S_N^{p^{\prime}p} [k_1,\varepsilon_1;\ldots;k_N,\varepsilon_N] & = & K_N^{p'p}[k_1, \varepsilon_1;\ldots;k_N,\varepsilon_N] (-\ps+m)
	\nonumber\\ && 
- e\sum_{i=1}^N K_{(N-1)}^{p',p+k_{i}}[k_1, \varepsilon_1;\ldots;\hat k_i,\hat \varepsilon_i; \ldots ;k_N,\varepsilon_N	] {\s \varepsilon_i}\, .
\label{SNreversed}
\ear
We equate the two expressions for $S_N$, multiply by $\left(p'^{2} + m^{2}\right)\left(p^2 + m^{2}\right)$, and 
then take $p,p'$ on-shell. This is sufficient to make the subleading terms drop out, since they all lack one of the two poles. 
Using \eqref{defcalK}, \eqref{defABC}   
we get the identity
\bear
 \slash p' (A_N \bone + B_{N\alpha\beta}\sigma^{\alpha\beta} -i C_N\gamma_{5})=
 (A_N \bone + B_{N\alpha\beta}\sigma^{\alpha\beta} -i C_N\gamma_{5})(-\slash p) \, ,
\ear
which by a comparison of the terms linear and cubic in Dirac matrices yields the identities
\bear
A_N(p+p')^{\alpha} &=& 2B_{N\,\,\nu}^{\,\, \,\,\alpha}(p-p')^{\nu}\, , \label{idAB1}\\
C_N(p-p')^{\alpha} &=& -2\widetilde B_{N\,\,\nu}^{\,\, \,\,\alpha}(p+p')^{\nu}\, . \label{idCB1}
\ear
Here we found it convenient to further introduce the dual tensor $\widetilde B_{N\mn} = \half \varepsilon_{\mu\nu\alpha\beta} B_N^\ab$.  
Contracting these with either $p_{\alpha}$ or $p'_{\alpha}$ we obtain scalar relations, 
\bear
A_N(m^2-p\cdot p') &=& 2p^\mu B_{N\mu\nu} p'^\nu \, ,\label{idAB2}\\
C_N(m^2+p\cdot p')&=& 2p^\mu  \widetilde B_{N\mu\nu}p'^\nu \, .\label{idCB2}
\ear
We see that knowledge of $B_{N\mu\nu}$ is sufficient to reconstruct the other coefficients. Note further that \eqref{idAB2} implies that $A_N=0$ for $p=p'$, while \eqref{idCB2} implies that $C_N=0$ for $p=-p'$. 

Aside from the identities presented here between coefficients at fixed $N$, there are some useful recurrence relations between coefficients for different numbers of photons: for brevity these are derived in Appendix \ref{app-recursion}.

\section{On-shell matrix elements for the dressed electron propagator}
\label{sec:matel}

We are now ready for our main task, namely the construction of the on-shell matrix elements corresponding to the fermion
line dressed with $N$ photons. We will treat these matrix elements in full generality - 
the fully polarised matrix elements, the spin-averaged polarised cross sections and the totally unpolarised cross section -
albeit only for the electron-electron case; the electron-positron, positron-electron and positron-positron cases
can be obtained from this by crossing as usual.

\subsection{Fully polarised spinor-line matrix element}
\label{sec:matferm}

Using \eqref{calMsspinfin}, \eqref{defABC} and formulas \eqref{DiracBilinearAC}, \eqref{DiracBilinearB} for the Dirac bilinears, we get the following
universal formulas for the polarised matrix elements $\mathcal{M}_{N\, s's}^{p'p}$, independently of the photon number and helicity
assignments (we recall that the fermion spins are defined with respect to the vectors $z^{\mu}$ and $r^{\mu}$ that enter the expressions for the coefficients above):
\begin{align}
\hspace{-3.5em}\mathcal{M}_{N\, ++}^{p'p} &= (-ie)^N \dfrac{ \mathcal{N}}{2m}  \bigg[ 
\alpha_1 \big( A_N - 4 d \cdot B_N \cdot d^\ast \big) -4 \, \alpha_2 \, z \cdot B_N \cdot d
+\dfrac{4}{m} \, \alpha_3 \, d \cdot B_N \cdot p'  + \alpha_4 \big( iC_N - \dfrac{2}{m} z \cdot B_N \cdot p' \big) \bigg],                                 \nonumber\\
\hspace{-3.5em}\mathcal{M}_{N\, +-}^{p'p} &= (-ie)^N \dfrac{ \mathcal{N}}{2m} \bigg[\!-\!\alpha_2^\ast \big(  A_N - 4 d \cdot B_N \cdot d^\ast \big) 
-4 \alpha_1 \, z \cdot B_N \cdot d  + \dfrac{4}{m} \, \alpha_4^\ast \, d \cdot B_N \cdot p'  - \alpha_3^\ast \big( i C_N - \dfrac{2}{m} z \cdot B_N \cdot p' \big) \bigg],  \nonumber\\
\hspace{-3.5em}\mathcal{M}_{N\, -+}^{p'p} &= (-ie)^N \dfrac{ \mathcal{N}}{2m} \bigg[ \alpha_2 \big( A_N + 4 d \cdot B_N \cdot d^\ast \big)+
4 \alpha_1 \, z \cdot B_N \cdot d^\ast - \dfrac{4}{m} \, \alpha_4 \, d^\ast \cdot B_N \cdot p' 
- \alpha_3 \big( i C_N + \dfrac{2}{m} z \cdot B_N \cdot p' \big) \bigg],                                   \nonumber\\
\hspace{-3.5em}\mathcal{M}_{N\, --}^{p'p} &= (-ie)^N \dfrac{ \mathcal{N}}{2m} \bigg[ 
\alpha_1 \big( A_N + 4 d \cdot B_N \cdot d^\ast \big) - 4 \, \alpha_2^\ast \, z \cdot B_N \cdot d^\ast
+ \dfrac{4}{m}\, \alpha_3^\ast \, d^\ast \cdot B_N \cdot p' - \alpha_4^\ast \big( i C_N + \dfrac{2}{m}\, z \cdot B_N \cdot p' \big) \bigg].       
\label{Mpolar}                          
\end{align}

\subsection{Summing over electron spins}
\label{subsec:spinsum}

For the construction of the spin-averaged cross section, we could either sum over the polarised matrix elements, 
\bear
\langle |\mathcal{M}^{p'p}_N|^2 \rangle &=& \frac{1}{2} \sum_{s,s'}|{\cal M}_{N s^{\prime}s}^{p^{\prime} p}|^2\, ,
\ear	
using our results \eqref{Mpolar}, or start directly from \eqref{calMsspinfin}, \eqref{defABC}. In either case the result reads, after simple algebra,
\bear
\hspace{-2em}\langle |\mathcal{M}^{p^{\prime} p}_N|^2 \rangle &=& \dfrac{e^{2N}}{2 m^2} \Big\{ \left(p'\cdot p + m^2\right) \left( |A_N|^2 + 2 B_N^{\alpha\beta}B^\ast_{N\alpha \beta} \right) + \left(p' \cdot p - m^2\right) |C_N|^2 \nonumber \\
\hspace{-2em}&& \hspace{30pt} -2 \Re \left[ 2 p' \cdot B_N \cdot p \ A_N^\ast - 4 p' \cdot B_N \cdot B_N^\ast \cdot p 
- 2p' \cdot \tilde B_{N}\cdot p \ C_N^\ast \right] \Big\}.
\label{suboptimal}
\ear
However, this expression can still be drastically simplified using the relations \eqref{idAB1}, \eqref{idCB1}. 
For this purpose, we start with rewriting (omitting now the subscript `$N$')
\bear
-2\Re \left[- 4 p' \cdot B \cdot B^\ast \cdot p\right]
=
2\Bigl[(p+p')\cdot B^{\ast}\cdot B\cdot (p+p') - (p-p')\cdot B^{\ast}\cdot B\cdot (p-p') \Bigr] \;. \quad
\ear 
From \eqref{idAB1} we have
\bear
4 (p-p')\cdot B^{\ast}\cdot B\cdot (p-p') = -(p+p')^2 |A|^2 = 2(m^2-p\cdot p')|A|^2 \;.
\ear
For the other term, we can use $B_{\mu\nu} = - \frac{1}{2}\varepsilon_{\mu\nu\alpha\beta}\tilde B^{\alpha\beta}$
to show
\bear
4 (p+p')\cdot B^{\ast}\cdot B\cdot (p+p') = 
4(p+p')\cdot \tilde B^{\ast}\cdot \tilde B \cdot (p+p')
- 2 (p+p')^2 B^{\ast}_{\alpha\beta}B^{\alpha\beta}
\ear
and then use \eqref{idCB1} to write
\bear
4(p+p')\cdot \tilde B^{\ast}\cdot \tilde B \cdot (p+p')
=
-(p-p')^2 |C|^2 
= 2(m^2+p\cdot p') |C|^2  \;.
\ear
Using these various relations in \eqref{suboptimal} together with \eqref{idAB2}, \eqref{idCB2} and their complex conjugates,
it can be transformed into the surprisingly
compact form that we wrote down already in the introduction, equation \eqref{optimal}:
\bear
\big\langle \big\pipe{\cal M}_N^{p'p}\big\pipe^2\big\rangle &=& e^{2N} \Bigl\lbrack \abs{A}^2 + 2 B_{}^\ab B_{\ab}^{\ast}
 - \abs{C}^2 
 \Bigr\rbrack .
 \label{optimalcopy}
\ear

\subsection{Summing over photon polarisations}
\label{subsec:sumpol}

Finally, summing over photon polarisations can be done using
\bear
\sum_{\lambda = \pm} \veps_{i\lambda}^{\mu}\veps_{i\lambda}^{\nu\ast} \to \eta^{\mn}\, .
\label{sumpol}
\ear
Here it should be remembered that, in the spinor-helicity formalism, this completeness relation 
normally would involve additional longitudinal terms on the right-hand-side, 
\bear\label{eq:polcomplete}
\sum_{\lambda = \pm} \veps_{i\lambda}^{\mu}\veps_{i\lambda}^{\nu\ast} 
= \eta^{\mu\nu} + { k_i^\mu q_i^\nu + k_i^\nu q_i^\mu \over k_i\cdot q_i }
\ear
(with $q_i$ the reference momentum of $\veps_i$)
but here they can be omitted due to the on-shell transversality of the coefficients $A_N,B_{N\mn},C_N$.

\section{The ``all +'' helicity case}
\label{sec:allplus}

Finally, let us have a closer look at the ``all +'' case. Here we can expect something special to happen because of the well-known
connections between helicity, duality, and supersymmetry \cite{thooft,dadda,brolee,dufish1,dufish2,51}. 
A background field consisting of photons with only ``+'' helicities is self dual, which in our present conventions means
(compare \eqref{idfdual})

\bear
\widetilde F_\mn = i F_\mn  \, .
\ear
This relation can equivalently be expressed as
\bear
F_\mn \sigma^\mn \cdot  (\Eins + \gamma_5) = 0\, ,
\label{susy}
\ear
which is also a step towards exhibiting the associated supersymmetry of the Dirac operator in a self-dual background 
\cite{thooft,dufish1,dufish2}. From the definition \eqref{defK} of the kernel $K_N^{p'p}$, and the fact that it reduces
to $D_N^{p'p}$ in the absence of the $F_\mn \sigma^\mn$ term, we conclude (e.g. by expanding $K_N^{p'p}$ in that term)
that in the all $+$ case
\bear
K_N^{p'p} \cdot  (\Eins + \gamma_5) = D_N^{p'p}  (\Eins + \gamma_5)\,,
\ear
leading to the following relations for the coefficient functions $A_N,B_{N\mn},C_N$, and $A_N^{\rm scal}$:
\bear
A_N-iC_N &=& A_N^{\rm scal} \label{AC} \, ,\\
\widetilde B_{N\mn} &=& i B_{N\mn}\, . \label{BB} 
\ear
Since $A_N = A_N^{\rm scal} +A_N^{\psi}$, the first of these relations can also be written as
\bear
A_N^{\psi} = i C_N\, .
\label{idAC}
\ear
Combining the second relation \eqref{BB} with \eqref{idAB2} and \eqref{idCB2}, we get a further relation 
\bear
(p\cdot p' - m^2) A_N = iC_N (p\cdot p' +m^2)\, ,
\label{idfurther}
\ear
and combining this equation with \eqref{idAC} enables us to express $A_N^\psi$ in terms of $A_N^{\rm scal}$:

\bear
A_N^{\psi} = \frac{p\cdot p' -m^2}{2m^2}A_N^{\rm scal}\, .
\label{idAA}
\ear
Turning our attention now to the spin-averaged cross section, further simplification follows from the fact that
in Minkowski space 
\bear
\widetilde{\widetilde B}_\mn = - B_\mn\, .
\ear
Combined with the self-duality relation \eqref{BB} this leads to the vanishing of $B^{\ast}_{N\alpha\beta}B_N^{\alpha\beta}$.
Thus \eqref{optimalcopy} now reduces to
\bear
\big\langle \big\pipe{\cal M}_N^{p'p}\big\pipe^2\rangle &=& e^{2N} \Bigl\lbrack \abs{A_N}^2 
 - \abs{C_{N}}^2
 \Bigr\rbrack ,
 \label{optimalallplus}
\ear
which furthermore by \eqref{idAC} and \eqref{idAA} can be expressed entirely in terms of the scalar coefficient $A_N^{\rm scal}$:
\bear
\big\langle \big\pipe{\cal M}_N^{p'p}\big\pipe^2\rangle &=& e^{2N} \frac{p\cdot p'}{m^2} \big\pipe A_N^{\rm scal}\big\pipe^2\, .
 \label{optimalallplusfin}
\ear
Therefore by \eqref{defANscal} we are getting a relation between the scalar and spinor cross sections, 
\bear
\big\langle \big\pipe{\cal M}_N^{p'p}\big\pipe^2\rangle &=& 
\frac{p\cdot p'}{m^2}  \big\pipe{\cal M}_{{\rm scal} N}^{p'p}\big\pipe^2\, .
 \label{scalarspinor}
\ear
This relation is, to the best of our knowledge, new. 
Note that it becomes equality for $p'=-p$,
\bear
\big\langle \big\pipe{\cal M}_N^{-p,p}\big\pipe^2\rangle &=& 
 \big\pipe{\cal M}_{{\rm scal} N}^{-p,p}\big\pipe^2\, ,
 \label{optimalallplusfinsusy}
\ear
a clear manifestation of the underlying supersymmetry. 

Note that equation \eqref{optimalallplusfin} cannot be used in the massless limit. Instead, we can conclude directly from \eqref{idfurther} that, for $m=0$, 
$A_N = iC_N$, leading to the vanishing of \eqref{optimalallplus} for the massless fermion line, and thus of the matrix elements
themselves, in agreement with \cite{DeCaus,StirlingSH, Ozeren, Badger2} (we also show this directly at the level of the amplitudes in appendix \ref{app-massless}).

\section{Compton scattering}
\label{sec:compton}

Let us now test our formalism on a recalculation of the Compton scattering amplitude and cross section in spinor QED. 
We would like to 
stress once more here that the underlying Feynman rules are not the standard Dirac ones, but second-order rules as explained in part I. 
As such they contain the diagrams that exist already in scalar QED, depicted in Fig. 1, and other ones involving at least one $\sigma^{\mu\nu}$ (spin) interaction, shown in
Fig.2. 

\begin{figure}[h]
\centering
    \includegraphics[width=0.8\textwidth]{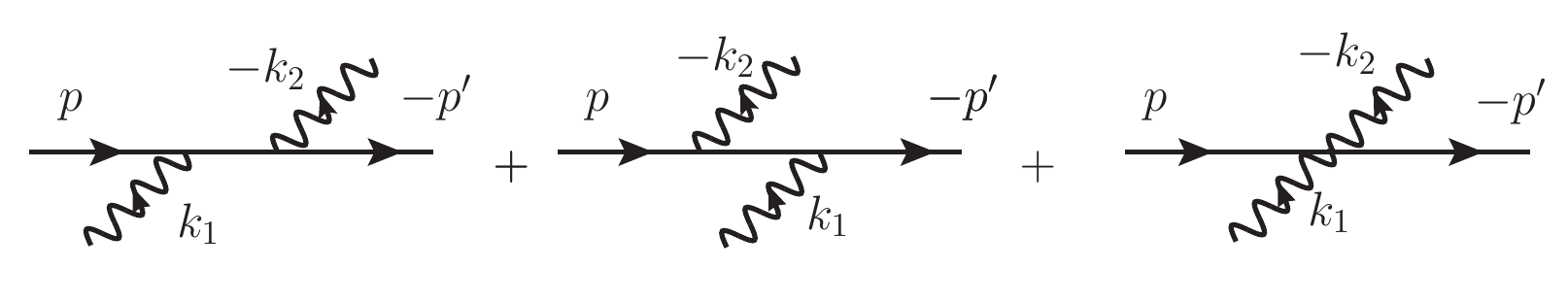}
  \caption{Scalar QED like contributions to the Compton scattering in spinor QED. We use $p,k_1$ as incoming and $-p', -k_2$ as outgoing momenta. }
  \label{2compton-scal}
\end{figure}

\begin{figure}[h]
\centering
    \includegraphics[width=0.8\textwidth]{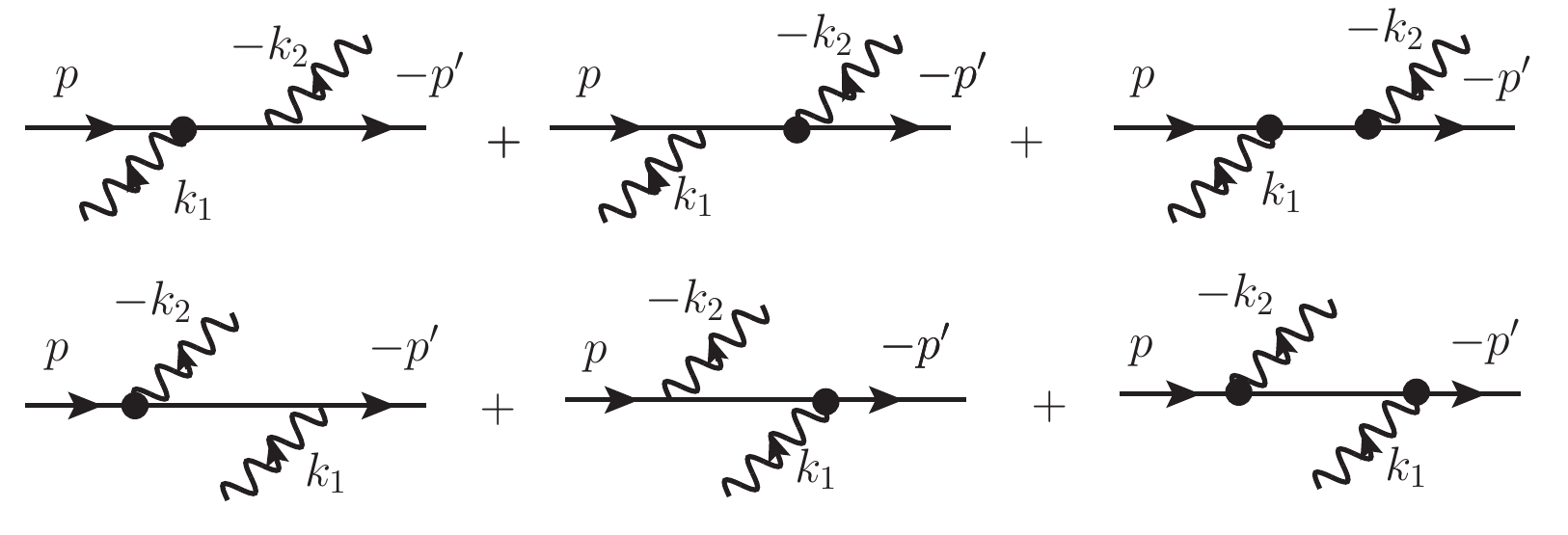}
  \caption{Extra diagrammatic contributions to the spinor case. The bullets represent the coupling with $\sigma^{\mu\nu}$ in the second order formalism, see Fig. 1 of I. We use $p,k_1$ as incoming and $-p', -k_2$ as outgoing momenta. }
  \label{2compton-spin}
\end{figure}

\subsection{The N=2 case}

For $N=2$, we already calculated the coefficient functions $A_2, B_{2\mn}, C_2$  off-shell in part I using the fermion line master formula, leading to (I 5.19), and we could use these expressions with on-shell conditions,
whereafter it is possible to rewrite them in terms of field-strength tensors. Instead, it is more convenient to follow our present approach
and get the on-shell coefficient functions using the spin-orbit decomposition from I and the $R$-representation. 
 
Using at first the same reference momenta of our scalar calculation above, \eqref{defrpm}, we find, using \eqref{K2decomp}, \eqref{barR1R2},
and (I 6.9), after simple algebra, the coefficients
\bear
A_2^{\rm scal} &=&  -2 \frac{r_1\cdot f_1\cdot f_2\cdot r_2}{r_1\cdot k_1 \, r_2\cdot k_2}
\, , \nonumber\\
A_2^{\psi} 
&=& -  \left( \frac{1}{2p'\cdot k_1} + \frac{1}{2p' \cdot k_2}  \right) W_{\eta = 0}
= - \frac{1}{2} \left( \frac{1}{2p'\cdot k_1} + \frac{1}{2p' \cdot k_2}  \right) \tr (f_1f_2)\, ,
\label{A2_spin}
\nonumber\\
B_2^{\mu\nu} &=& 
-\frac{1}{2} \frac{ r_1 \cdot f_1 \cdot k_2 f_2^{\mu\nu} + r_2 \cdot f_2 \cdot k_1 f_1^{\mu\nu}  }{r_{1} \cdot k_1 r_{2} \cdot k_2}
+ \frac{1}{2} \left( \frac{1}{2p'\cdot k_1} - \frac{1}{2p' \cdot k_2}  \right)  ( [f_1 , f_2 ])^{\mu\nu}
\nonumber\\ 
C_2 
&=& - \frac{1}{4} \left( \frac{1}{2p'\cdot k_1} + \frac{1}{2p' \cdot k_2}  \right) \epsilon_{\alpha\beta\gamma\delta} f_1^{\alpha\beta} f_2^{\gamma\delta} = \frac{1}{2} \left( \frac{1}{2p'\cdot k_1} + \frac{1}{2p' \cdot k_2}  \right) \text{tr}(f_1 \widetilde{f}_2)\, .
\nonumber\\
\label{ABCfinal}
\ear
At this stage it becomes clear that the part of each of the reference vectors \eqref{defrpm} proportional to their photon's momentum drops out of the scalar products in each coefficient. As such, we could at this stage replace the $r_{i}$ by $p^{\prime}$ without losing information (since this does not invalidate our removal of the more complicated term in $\bar{R}_{2}$) -- the same is true for scalar QED as is clear from (\ref{Mscal+-fin}) and its correspondence with $A_{2}^{\textrm{scal}}$.

Below we shall make use of both representations of the coefficients. In general it is useful to keep the $r_{i}$ null vectors, for use with the spinor helicity representation (section \ref{phopol}). However, it turns out that for calculating the amplitude in a particular reference frame, it is more convenient to take advantage of the freedom to use $r_{i} = p^{\prime}$, since the explicit construction of the spinors $\pipe r_{i}\rangle$, $\pipe r_{i} ]$ etc, is complicated by their dependence on the photon momenta. On the other hand, at the level of the cross section, spinor products turn into ($4$-vector) scalar products and here the vector form of the $r_{i}$ can be applied directly.
 
\subsection{The functions $A_2, B_{2\ab}, C_2$ with fixed helicities}
We begin by providing the coefficients in (\ref{ABCfinal}) in an arbitrary Lorentz frame for the construction of the fully polarised amplitudes in \eqref{Mpolar} for $N=2$. Since also the vectors $d^{\mu}$ and $d^{\ast \mu}$ are defined for an arbitrary frame 
(likewise $z^{\mu}$ and $r^{\mu}$ can be chosen with any direction in this frame), this achieves a representation of the polarised amplitudes valid for a general system of reference.

We will denote the polarised amplitudes by $\mathcal{M}_{2\, s's}^{h_1h_2}$, suppressing the superscript `$p'p$' and as above denoting the
helicity of photon $i$ by $h_i$. For the calculation of $\mathcal{M}_{2\, s's}^{++}$, we require the coefficients
\begin{align}
	A_{2}^{\textrm{scal}++} &= -2 [12]^2\frac{m^{2}}{\lambda_{1}\lambda_{2}} \, ,\nonumber \\
	A_{2}^{\psi ++} &= \frac{1}{2}[12]^{2}\frac{\lambda_{1} + \lambda_{2}}{\lambda_{1}\lambda_{2}}\, ,\nonumber \\
	B_{2}^{\mu\nu ++} &= \frac{[12]}{8\lambda_{1}\lambda_{2}}\Big[(\lambda_{2} - \lambda_{1}) [1\pipe [\gamma^{\mu}, \gamma^{\nu}]\pipe 2]- \lambda_{1}\frac{\langle r_{1}2\rangle}{\langle r_{1}1\rangle}[2 \pipe [\gamma^{\mu}, \gamma^{\nu}]\pipe 2]+ \lambda_{2}\frac{\langle r_{2}1\rangle}{\langle r_{2}2\rangle}[1 \pipe [\gamma^{\mu}, \gamma^{\nu}]\pipe 1] \Big]
	\nonumber\\
	&=
	\frac{[12]}{8\lambda_{1}\lambda_{2}}\Big[(\lambda_{2} - \lambda_{1}) [1\pipe [\gamma^{\mu}, \gamma^{\nu}]\pipe 2]
	-\lbrack1\vert\slash p'\vert2\rangle [2 \pipe [\gamma^{\mu}, \gamma^{\nu}]\pipe 2]
	+ \lbrack2\vert\slash p'\vert1\rangle
	[1 \pipe [\gamma^{\mu}, \gamma^{\nu}]\pipe 1] \Big]
	\nonumber\\
	&=
	\frac{[12]^2}{8\lambda_{1}\lambda_{2}}
	\Big[
	 \langle 1\pipe \slash p' [\gamma^{\mu}, \gamma^{\nu}]\pipe 1] +  \langle 2\pipe \slash p' [\gamma^{\mu}, \gamma^{\nu}]\pipe 2]
	 \Big]
	\, ,\nonumber \\
	C_{2}^{++} &= -\frac{i}{2}[12]^{2} \frac{\lambda_{1} + \lambda_{2}}{\lambda_{1}\lambda_{2}}
	\label{eqCoefs++}
\end{align}
(recall that $\lambda_i = 2 p' \cdot k_i$). With these coefficients 
one readily verifies the general results of sections \ref{sec:relations} (equations (\ref{idAB2}) and (\ref{idCB2})) and \ref{sec:allplus} (equations (\ref{idAC})--(\ref{idAA})). In particular, we shall use the important result $C_{2}^{++} = -iA_{2}^{\psi ++}$ below. 
For the amplitudes $\mathcal{M}_{2\, s's}^{+-}$ we find
\begin{align}
	A_{2}^{\textrm{scal}+-} &= -2 \frac{\langle r_{1}2 \rangle[1r_{2}]}{\langle r_{1}1\rangle [2r_{2}]}
	= - 2
\frac
{\lbrack 1\vert \slash p' \vert 2 \rangle^2}
{\lambda_1\lambda_2}
 \, ,\nonumber \\
	A_{2}^{\psi +-} &= 0\, ,\nonumber \\
	B_{2}^{\mu\nu +-} &= \frac{1}{8\lambda_{1}\lambda_{2}}\Big[ \lambda_{1}\frac{\langle r_{1}2\rangle [21]}{\langle r_{1}1\rangle}\langle 2 \pipe [\gamma^{\mu}, \gamma^{\nu}]\pipe 2\rangle+ \lambda_{2}\frac{[ r_{2}1]\langle 12\rangle }{[r_{2}2]}[1 \pipe [\gamma^{\mu}, \gamma^{\nu}]\pipe 1] \Big]
	\nonumber\\
	&= \frac{[1\vert\slash p'\vert 2\rangle}{8\lambda_{1}\lambda_{2}}
	\Big[
\langle 12 \rangle \lbrack 1 \pipe [\gamma^{\mu}, \gamma^{\nu}]\pipe 1\rbrack
		-[12]\langle 2 \pipe [\gamma^{\mu}, \gamma^{\nu}]\pipe 2\rangle
		\Big]
	\, ,\nonumber \\
	C_{2}^{+-} &= 0\, .
	\label{eqCoefs+-}
\end{align}
Again, these results satisfy the relations given in section \ref{sec:relations}. These coefficients can be substituted directly into \eqref{Mpolar} to find the fully polarised amplitudes in any reference frame (we detail the determination of the coefficients $\alpha_{i}$ in the following subsection). 

\subsection{The fully polarised amplitudes}

We now further specialise to the centre-of-mass frame, where
\begin{align}
p_{cm}^{\prime \mu} &= - E_e \, ( 1, \beta \sin \theta, 0, \beta \cos \theta )\,, &
p_{cm}^{\mu} &= E_e \, ( 1,0,0,\beta )\,, \nonumber \\ 
k_{1 \,cm}^{\mu} &= E_\gamma \, ( 1, 0, 0, -1 )\,, &
k_{2 \, cm}^{\mu} & = - E_\gamma \, ( 1, -\sin \theta, 0, -\cos \theta )\,,
\end{align}
with
\begin{eqnarray}
\beta = \dfrac{E_\gamma}{E_e}, \qquad E^2_e - E_\gamma^2 = m^2.
\end{eqnarray}
Measuring the particle spins along their direction of motion (``helicity basis''), we can use the general formulas derived in appendix \ref{app-bispecial} with
$E=E'=E_e$ and $\absp=\abspr = E_\gamma$. From \eqref{app-zstandard}, \eqref{app-zhel} and
\eqref{app-dhel} we get
\begin{align}
z_{cm}^{\mu} &= \dfrac{E_e}{m} \, ( \beta, \sin \theta, 0, \cos \theta )\,, &
r_{cm}^{\mu} &= \dfrac{E_e}{m} \, ( \beta, 0, 0, 1 )\,, \nonumber \\
d_{cm}^{\mu} &= \dfrac{1}{2} \, ( 0, \cos \theta, -i, -\sin \theta )
\end{align}
and the coefficients $\alpha_i$ can be read off from \eqref{app-alphahelicity}:
\bear
\alpha_1 &=& \half  (1+\cos\theta) = \cos^2\frac{\theta}{2}
\nonumber\\
\alpha_2 &=& - \frac{E_e}{2m} \sin\theta = -  \frac{E_e}{m} \sin\frac{\theta}{2}\cos\frac{\theta}{2}
\nonumber\\ 
\alpha_3 &=&  - \frac{E_\gamma}{2m}  \sin\theta = -  \frac{E_\gamma}{m} \sin\frac{\theta}{2}\cos\frac{\theta}{2}
\nonumber\\
\alpha_4 &=& 0 \;.
\label{alphahelicity}
\ear
When written in terms of the Mandelstam variables
\bear
s &=& - (p+k_1)^2 = -(p'+k_2)^2 = m^2-\lambda_2 = (E_e+E_\gamma)^2\,, \nonumber \\
u &=& - (p+k_2)^2 = - (p'+k_1)^2 = m^2-\lambda_1 = (E_e-E_\gamma)^2-4E_\gamma^2 \cos^2 \dfrac{\theta}{2}\,, \nonumber \\
t &=& -2 k_1 \cdot k_2  = -2 k_1 \cdot k_2 = - 4 E_\gamma^2 \sin^2 \dfrac{\theta}{2}\,.
\ear 
the non-zero coefficients turn into
\bear
\mathcal{N}\alpha_1 &=& 2m \dfrac{\sqrt{m^4-su}}{s-m^2}\,, \nonumber\\
\mathcal{N}\alpha_2 &=& - \dfrac{s+m^2}{s-m^2} \sqrt{-t}\,, \nonumber \\
\mathcal{N}\alpha_3 &=& -\sqrt{-t}\, . 
\label{alphamandel}
\ear
We shall again provide the amplitudes $\mathcal{M}_{2\, s's}^{++}$ and $\mathcal{M}_{2\, s's}^{+-}$ using \eqref{ABCfinal} and (\ref{Mpolar}) in this frame. Since in the coefficients $A_2, B_{2 \mu\nu}$ and $C_2$ we represent the field strength tensors in terms of spinor helicity variables, we need the explicit components of the spinors related to the vector $k_{1\,cm}$ and $k_{2\,cm}$,
\begin{align}
|1\rangle^{\dot{a}} &= \sqrt{2E_\gamma} \begin{pmatrix}
0 \\ 1
\end{pmatrix}, & |1]_a &= \sqrt{2E_\gamma} \begin{pmatrix}
-1 \\ 0
\end{pmatrix}\,, \nonumber \\
|2\rangle^{\dot{a}} &= - \sqrt{2 E_\gamma} \begin{pmatrix}
\sin \frac{\theta}{2} \\ - \cos \frac{\theta}{2}
\end{pmatrix}, & |2]_a &= \sqrt{2 E_\gamma} \begin{pmatrix}
\cos \frac{\theta}{2} \\ \sin \frac{\theta}{2}
\end{pmatrix}\,.
\end{align}
For the actual calculation of the coefficients, however, it is now more convenient to replace the $r_{i} \rightarrow p^{\prime}$ to avoid having to determine the explicit form of the spinors for the reference vectors in this frame. Doing this in \eqref{ABCfinal}, we obtain,
\begin{align}
A_2^{++} &= \dfrac{t (4m^2-t)}{2(s-m^2)(u-m^2)}\,, \quad &
C_2^{++} &= i \dfrac{t^2}{2(s-m^2)(u-m^2)}\,,  \\
d \cdot B_2^{++} \cdot p' &= m^2 \, \dfrac{t \sqrt{-t (m^4-su) } }{4(s-m^2)^2 (u-m^2)}\,, \quad &
z \cdot B_2^{++} \cdot p' &= - \dfrac{m t^2 (s+m^2)}{4(s-m^2)^2(u-m^2)}\,, 
\end{align}
and we can express the remaining scalars in terms of these:
\bear
 z \cdot B_2^{++} \cdot d &=& \frac{1}{m}d \cdot B_2^{++} \cdot p'\, ,\\
 \, d \cdot B_2^{++} \cdot d^\ast &=& \frac{1}{2m}z\cdot B_2^{++} \cdot p'\,,\\ 
   \; z \cdot B_2^{++} \cdot d^\ast &=& -\frac{1}{m}d \cdot B_2^{++} \cdot p'\, ,\\
    d^{\ast} \cdot B_2^{++} \cdot p^{\prime} &=&  d \cdot B_2^{++} \cdot p'\,.
\ear

Using \eqref{Mpolar}, and the above equations, the fully polarised $\mathcal{M}_{2\, s's}^{++}$ amplitudes read as,
\begin{align}
\mathcal{M}_{2\, ++}^{++} &= - 2 e^2 m^2 \dfrac{t \sqrt{m^4-su}}{(s-m^2)^2 (u-m^2)}\,, \quad &
\mathcal{M}_{2\, +-}^{++} &= - 2 e^2 m^3 \dfrac{t\sqrt{-t}}{(s-m^2)^2(u-m^2)}\,, \nonumber \\
\mathcal{M}_{2\, -+}^{++} &= 2 e^2 m \dfrac{s t \sqrt{-t}}{(s-m^2)^2(u-m^2)}\,, \quad &
\mathcal{M}_{2\, --}^{++} &= - 2 e^2 m^2 \dfrac{t \sqrt{m^4-su}}{(s-m^2)^2 (u-m^2)}\,.
\end{align}
Likewise for the $\mathcal{M}_{2\, s's}^{+-}$ amplitudes we find 
\begin{align}
\hspace{-3em}A_2^{+-} &= \dfrac{-2(m^4 - su)}{(s-m^2)(u-m^2)}\,, \quad &
C_2^{+-} &= 0\,,  \\
d \cdot B_2^{+-} \cdot p' &= - \dfrac{(m^4-su)^{\frac{3}{2}} \sqrt{-t} }{4 (s-m^2)^2 (u-m^2)} \,, \quad &
d^\ast \cdot B_2^{+-} \cdot p' &= - \dfrac{\sqrt{-t(m^4-su)} \left[ (s-m^2)^2 + m^2 t \right] }{4 (s-m^2)^2 (u-m^2)}, \\
\hspace{-3em}z \cdot B_2^{+-} \cdot p' &= \dfrac{m t (m^4-su)}{2(s-m^2)^2 (u-m^2)}\,, \quad &
z \cdot B_2^{+-} \cdot d^\ast &= - \dfrac{\sqrt{-t(m^4-su)} \left[ (s-m^2)^2 - m^2 t \right] }{4 m (s-m^2)^2 (u-m^2)},
\end{align}
along with 
\bear
 z \cdot B_2^{+-} \cdot d &=& \frac{1}{m}d \cdot B_2^{+-} \cdot p'\, ,\\
 \, d \cdot B_2^{++} \cdot d^\ast &=& \frac{1}{2m}z\cdot B_2^{+-} \cdot p'\, ,
\ear
which lead to 
\begin{align}
\mathcal{M}_{2\, ++}^{+-} &= 2e^2 \dfrac{(m^4-su)^{\frac{3}{2}}}{(s-m^2)^2(u-m^2)} \,, \quad &
\mathcal{M}_{2\, +-}^{+-} &= 2e^2 m \dfrac{\sqrt{-t}(m^4-su)}{(s-m^2)^2(u-m^2)} \,,\nonumber\\
\mathcal{M}_{2\, -+}^{+-} &= -2e^2 m \dfrac{\sqrt{-t}(m^4-su)}{(s-m^2)^2(u-m^2)} \,,\quad &
\mathcal{M}_{2\, --}^{+-} &= 2e^2 \dfrac{\sqrt{m^4-su} \left[ (s-m^2)^2+m^2t \right] }{(s-m^2)^2(u-m^2)} \,. 
\end{align}
All these matrix elements agree, up to signs due to differing conventions, with the results presented 
by Denner and Dittmaier \cite{Denner}
(to be precise, our $\mathcal{M}_{2\, s's}^{h_1h_2}$ corresponds to their $\mathcal{M}_0(s,-h_1,s',h_2)$,
with a relative sign $(-1)^{s+s'+h_1+h_2+1}$). 
For historical discussion and derivation of Compton scattering from density matrix and operator techniques, and using the Stokes parameters for the description of polarisation, we refer the reader to \cite{fano,mcmaster}.

\subsection{Polarised photons, unpolarised electrons}
Moving on to the cross section, we can complete the electron spin sums whilst leaving the photon helicities arbitrary. For this calculation, it is useful to use (\ref{eqCoefs++}) and (\ref{eqCoefs+-}) as they stand (rather than replacing the $r_{1,2}$ by $p'$),
since upon multiplication of the coefficients with their complex conjugates, the spinor products are turned into $4$-vector scalar products. We begin with the amplitudes for the $++$ helicity assignment. Noting that 
\bear
2k_1\cdot k_2 = - (\lambda_1+\lambda_2)\, ,
\label{idktolambda}
\ear
we find, after simple algebra,
\bear
\abs{A _2^{++}}^2 &=& 
\frac{1}{\lambda_1^2\lambda_2^2} \, (k_1\cdot k_2)^2 (\lambda_1+\lambda_2+4p'^2)^2 
=
\frac{(\lambda_1+\lambda_2)^2}{4\lambda_1^2\lambda_2^2} \, (\lambda_1+\lambda_2-4m^2)^2 \, ,
 \nonumber\\
 B_2^{\mu\nu ++}(B_{2\mu\nu}^{++})^{\ast} &=& 0 \, ,\nonumber\\
\abs{C _2^{++}}^2 &=& \frac{1} {\lambda_1^2\lambda_2^2} \, (k_1\cdot k_2)^2 (\lambda_1+\lambda_2)^2 
=
\frac{(\lambda_1+\lambda_2)^4}{4\lambda_1^2\lambda_2^2} \,, 
\label{A2B2C2onshell++}
\ear
so that (\ref{optimalcopy}) provides
\bear
\langle \abs{{\cal M}_2^{++}}^2\rangle &=&  e^4 \frac{(\lambda_1+\lambda_2)^2}{4\lambda_1^2\lambda_2^2} \, \Bigl\lbrack (\lambda_1+\lambda_2-4m^2)^2
- (\lambda_1+\lambda_2)^2\Bigr\rbrack \, .
\ear
Note that we could have obtained this result more easily using \eqref{Mscal++fin} and \eqref{scalarspinor}. 

Similarly for the $+-$ helicity amplitude we find
\bear
\abs{A _2^{+-}}^2  &=& 
\frac{4}{\lambda_1^2\lambda_2^2} \, (2 (k_{1}\cdot k_{2})p'^2-\lambda_1\lambda_2)^2 = \frac{4}{\lambda_1^2\lambda_2^2} \, 
\Bigl\lbrack m^2(\lambda_1+\lambda_2) - \lambda_1\lambda_2\Bigr\rbrack^2\, ,
 \nonumber\\
 B_2^{\mu\nu +-}(B_{2\mu\nu}^{+-})^{\ast} &=& \frac{4}{\lambda_1^2\lambda_2^2} \,(k_{1}\cdot k_{2})^2 (2(k_{1}\cdot k_{2}) p'^2-\lambda_1\lambda_2)
 =
 \frac{(\lambda_1+\lambda_2)^2}{\lambda_1^2\lambda_2^2} 
\Bigl\lbrack m^2(\lambda_1+\lambda_2)- \lambda_1\lambda_2\Bigr\rbrack\, ,
 \nonumber\\
\abs{C _2^{+-}}^2  &=& 0 \, ,
\label{A2B2C2onshell+-}
\ear
giving in total
\bear
\langle \abs{{\cal M}_2^{+-}}^2\rangle &=& 
 e^4 \frac{2}{\lambda_1^2\lambda_2^2} \, 
\Bigl\lbrack m^2(\lambda_1+\lambda_2) - \lambda_1\lambda_2\Bigr\rbrack
\Bigl\lbrack \lambda_1^2 + \lambda_2^2 + 2 m^2(\lambda_1+\lambda_2)\Bigr\rbrack\, .
\ear
Writing these results in terms of the usual Mandelstam variables introduced above we arrive at
\begin{align}
\begin{split}
\langle \abs{{\cal M}_2^{++}}^2\rangle &= 4e^4 \frac{m^2 t^2(m^2-\frac{t}{2})}{(m^2-u)^2(m^2-s)^2}\,, 
\\
\langle \abs{{\cal M}_2^{+-}}^2\rangle &=
4 e^4 \frac{(us-m^4)(us-\frac{t^2}{2}-m^4)}{(m^2-u)^2(m^2-s)^2}\, .
\end{split}
\label{Tpolarised}
\end{align}
In the massless limit, this gives the well-known results
\begin{align}
\begin{split}
\langle \abs{{\cal M}_2^{++}}^2\rangle &= 0 \, ,\\
\langle \abs{{\cal M}_2^{+-}}^2\rangle &= 
-2 e^4
\Bigl( \frac{s}{u} + \frac{u}{s} \Bigr) \, .
\end{split}
\label{Tpolarisedmassless}
\end{align}
Note that, in our conventions, the $++$ amplitude is the helicity violating one, correctly vanishing in the massless case. In the present formalism, it is clear from (\ref{A2B2C2onshell++}) that this comes about by a cancellation between the $A$ and $C$ terms. We discuss some aspects of the massless amplitudes in greater detail in appendix \ref{app-massless}. 

\subsection{The unpolarised Compton cross section}
Finally, let us also sum over photon polarisations to write down the unpolarised cross section:
\bear
\frac{1}{2} \sum_{\lambda,\lambda'}
\big \langle \big\pipe{\cal M}_2^{\lambda \lambda'}\big\pipe^2\big\rangle 
&=&
\big \langle \abs{{\cal M}_2^{++}}^2\big \rangle
+
\big \langle \abs{{\cal M}_2^{+-}}^2\big \rangle 
\nonumber\\
&=&
2e^4
\biggl\lbrace
-\frac{\lambda_1}{\lambda_2} -\frac{\lambda_2}{\lambda_1} - 4m^2 \Bigl(\frac{1}{\lambda_1} + \frac{1}{\lambda_2} \Bigr) 
+4m^4 \Bigl(\frac{1}{\lambda_1} + \frac{1}{\lambda_2} \Bigr)^2 
\biggr\rbrace\, .
\nonumber\\
\label{comptonunpol}
\ear
This is again in agreement, of course, with the standard textbook result, see, for example, eq. (5.87) in \cite{peskinschroeder-book}. 

\section{Conclusions and Outlook}
\label{sec:conc}

In the second part of this series of papers, we have worked out the simplifications that can be achieved using the new, first-quantised worldline representation of the fermion propagator developed in part 1 for the photon-dressed fermion line, when both the fermion legs as well as all
photons are taken on-shell (although it should be emphasised that
many of the formulas presented here would still be valid if only the fermions were taken on-shell,
not the photons). For general kinematics and helicity assignments, we have found three types of simplifications:

\begin{enumerate}

\item
Most importantly, the ``subleading'' contributions to the dressed propagator can be omitted.

\item
The coefficient functions $A_N,B_{N\mn},C_N$ become manifestly transversal and can be rewritten in terms of photon field-strength
tensors in a systematic way. 

\item
There exist $N$-independent relations between those coefficient functions. As a consequence, all on-shell information is contained in the coefficient $B_{N\mu\nu}$. 

\end{enumerate}

Further simplifications have been exhibited for special kinematics ($p' = \pm p$) or helicity assignments (the ``all +'' case).
We have also derived recursion formulas for the functions $A_N,B_{N\mn},C_N$ that may become useful as an alternative to the
direct calculation methods of I.

On a technical level, we have provided a number of useful relations involving the fixed-helicity photon field-strength tensors,
and we have derived explicit formulas for all the relevant Dirac bilinears, for arbitray on-shell momenta $p,p'$ and 
spin vectors $r,z$. These formulas have allowed us to give expressions for the fully polarised amplitudes valid in an arbitrary Lorentz frame in terms of a canonical set of numerical coefficients and reference vectors defined in that frame. For the fixed-helicity but spin-summed cross sections we have found that the result can be derived purely from the squared moduli of the coefficients.

Applying this machinery to the linear Compton scattering process we have found it to lead to significant simplifications
over the standard calculation in the first-order Dirac formalism\footnote{R. Stora once told one of the authors (C.S.) that V. Weisskopf was so dissatisfied with the standard textbook calculation of the
Compton cross section that he asked not only him, but several others of his PhD students to look for a better way to get the simple
final result.}. The reasons are easy to pinpoint: first, the encoding of the Dirac algebra structure in the Grassmann path integral replaces the  ``partially antisymmetric'' (in the Lorentz indices) Dirac matrices 
by the ``fully antisymmetric'' vectors $\eta^{\mu}$, and therefore in $D=4$ leads effectively to an early projection on the Clifford basis, avoiding the appearance of products of more than four Dirac matrices. 
We find it curious that, although this representation including the
``symbol map'' have in principle been known for decades \cite{fradkin,fragit}, 
this attractive feature has apparently never been noted.
This also implies an early disentanglement of the photon polarisations and fermion spins, allowing them to be treated completely independently. In particular it makes it possible to perform the
average over the latter without having to fix the photon number or helicity assignments. 

Thus we expect the formalism to be very well-suited for future calculations of non-linear Compton scattering
\cite{King1,Boca1,Heinzl1,Seipt1,Seipt2,Boca2,King2}, including at higher orders, as well as its crossed process, multiple photon
Bremsstrahlung 
(see, for example, \cite{Haug, Gupta, Majumdae, Nadzhafov, DeCausmaeckerI, BerendsII, BerendsIV, BerendsVI}) 
and electron-positron annihilation into several photons \cite{acfpc,eidkur,berkle,lee-epannihilation}.
This will, of course, require a computer algebra implementation, which is in progress. 

\begin{figure}[h]
\centering
    \includegraphics[width=0.5\textwidth]{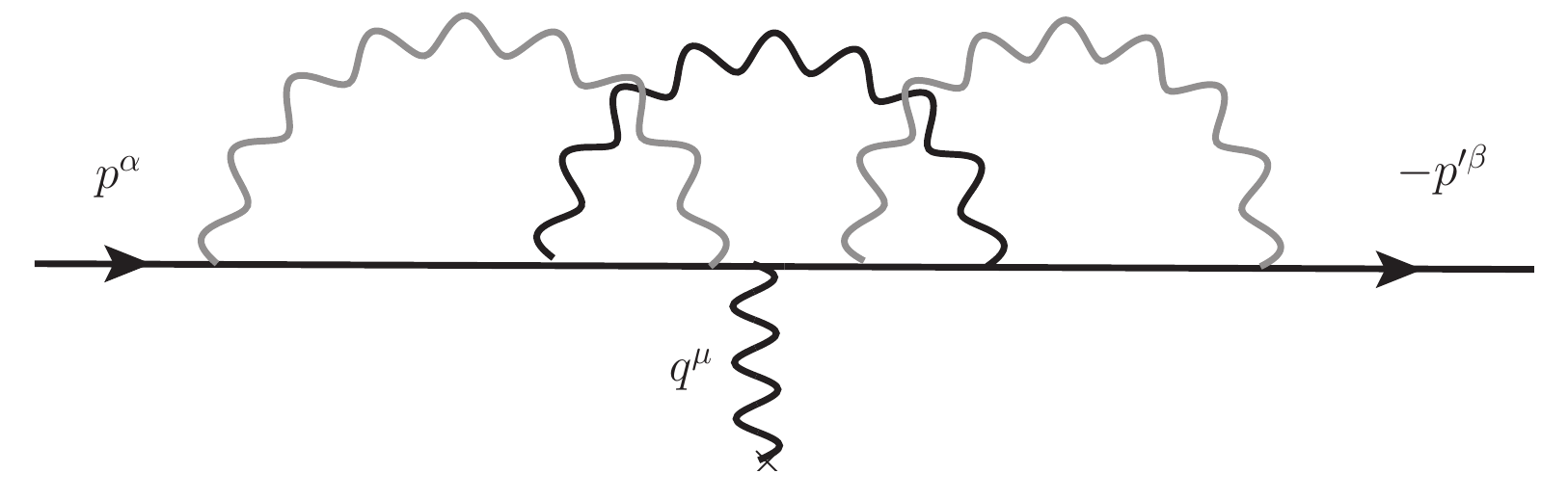}
  \caption{Single fermion line contributions to the QED $g-2$ factor. }
  \label{fig-g2}
\end{figure}

Another very natural application would be to multiloop contributions to the QED anomalous magnetic moment of the type
shown in Fig. \ref{fig-g2}.
A version of this formalism taylored to this specific purpose ($p'=-p$, one photon taken at low energy,
the remaining ones taken off-shell and sewn off in pairs) will be presented elsewhere. As an additional attractive feature,
the formalism makes it relatively easy to extract the form factor $F_2$.

In the forthcoming third part of this series, we extend the formalism to the construction of the dressed electron propagator in a constant external field,
a case where the absence of a practicable extension of the worldline formalism to open fermion lines had been particularly felt 
(partial results of the third part have already been published in \cite{113}). 

However, there are many other possible generalisations, let us mention here only a few:
\begin{itemize}

\item
Replacing the photons by other bosons coupling to the fermion line by Yukawa or axial couplings (for the closed-loop case
suitable formalisms have been available for decades \cite{12,16,dhogag1,dhogag2,29,33}). 

\item
Dressing the electron line with gravitons, which would generalise previous work on closed 
scalar and fermion loops \cite{Bastianelli:2004zp,Bastianelli:2007jv,87} as well as on open scalar lines~\cite{125}. 
This is of current interest considering the recent application of the worldline formalism to classical black-hole scattering
\cite{plefka,jmps1,jmps2}.

\item
Dressing a quark line with gluons, where the color degrees of freedom can be treated very similarly to the spin ones using Grassmann variables
\cite{Balachandran:1976ya,Barducci:1976xq,Bastianelli:2013pta,Ahmadiniaz:2015xoa, JO1, JO2}.
\item
Going to finite temperature (see, e.g.,  \cite{sato:therm,mckeon:therm,venwir}). 

\end{itemize}

As a final remark, let us emphasise that although here we have focused on the implementation of the formalism in terms of path integrals, nowadays
often simply called the ``worldline formalism,'' we could also have worked more directly with the Feynman diagrams of the second-order formalism.
However, such an approach would obscure some of the important advantages of the formalism, such as its ability to combine Feynman diagrams
that differ only by the position of the photon legs along the fermion line, and the flexibility provided by IBP in the proper-time parameters. 

\acknowledgments

We thank P. Cvitanovi\'{c}, G.V. Dunne and A. Ilderton for useful conversations and correspondence. 
CS and JPE thank CONACYT for support through project Ciencias Basicas 2014 No. 242461. JPE further acknowledges financial support from UMSNH through CiC project \#483224-2019. VMBG received support from PRODEP project 511-6/19-4990 for part of this work. 

\appendix

\section{Conventions}
\label{app-conv}
On the side of the worldline formalism, we work in Euclidean space with metric $(+ + + \, +)$, and use
Dirac matrices fulfilling $\lbrace \gamma^{\mu},\gamma^{\nu}\rbrace = -2 \delta^{\mu\nu}$.
On the field theory side, we Wick rotate to Minkowski space with metric $\eta_{\mu\nu} = \textrm{diag}(- + + \, +)$, 
and use $\lbrace \gamma^{\mu},\gamma^{\nu}\rbrace = - 2 \eta^{\mu\nu}$. We further 
define $\varepsilon^{0123} = + 1$ and $\gamma_5 = i\gamma^0\gamma^1\gamma^2\gamma^3$. 
These Minkowski space conventions coincide with the textbook of Srednicki \cite{srednicki-book} except for the sign of the electric charge, and the
fact that we use all ingoing momenta in Feynman diagrams instead of all outgoing ones. 

\section{Special cases of Dirac bilinears}
\label{app-bispecial}

In this appendix, which supplements section \ref{sec:bilinears},
we compute the coefficients $\alpha_1, \ldots, \alpha_4$ explicitly for the 
electron-electron case and for the two most common
choices of the spin-axis vectors $r^\mu$ and $z^\mu$, 
corresponding to a projection on the electron's direction of motion (the helicity basis) or on the direction corresponding to the $z$-axis in its rest frame. 
As an overall convention, we take the ingoing electron to move along the $z$ axis and the
outgoing one in the $x$-$z$ plane. Thus we can write their four-vectors as
\bear
p^\mu = (E,0,0,\absp) \,, \quad -{p'}^\mu = (E',\abspr\sin \theta,0, \abspr\cos \theta) \;.
\label{convp}
\ear

\subsection{Standard basis}

If we use the most common definition of the electron polarizations, the vectors $r^\mu$ and $z^\mu$ 
are taken as $(0,0,0,1)$ in the respective rest frames, and then subjected to the same Lorentz boost
that brings their electron up to speed. With our convention \eqref{convp} this results in
\begin{align}
r^\mu &= \frac{1}{m} (\absp,0,0,E)  \label{app-r}\\
z^\mu &=  \frac{1}{m} \bigl(\abspr \cos\theta, (E'-m) \cos\theta \sin\theta,0,E'\cos^2\theta + m \sin^2\theta\bigr) .
\label{app-zstandard}
\end{align}
Finally, the null vector \eqref{Def_vectord} 
\bear
d^\mu &= \dfrac{1}{4m} \overline{u}_+(-p') \gamma^\mu \gamma_5 u_-(-p')
\ear
must be constructed using the corresponding spinorial Lorentz boost. This gives, after a lengthy but
simple calculation,
\bear
d^\mu = \frac{1}{2m} \bigl(\abspr\sin\theta, E'\sin^2\theta + m\cos^2\theta,-im, (E'-m)\sin\theta\cos\theta\bigr) \;.
\ear
This is all we need for the evaluation of \eqref{DiracScalars}. One finds, after substantial cancellations,
\bear
4m^2 \alpha_1 &=& (E+m)(E'+m) - 2\absp\abspr \cos \theta + (E-m)(E'-m) \cos^2\theta 
\nonumber\\
4m^2 \alpha_2 &=& - \absp\abspr \sin \theta + (E-m)(E'-m) \sin\theta\cos\theta
\nonumber\\ 
4m^2\alpha_3 &=& 
\absp (E'-m)\sin \theta \cos\theta - \abspr (E+m)\sin\theta \nonumber\\
4m^2\alpha_4 &=& 
-\absp\bigl\lbrack (E'+m) + (E'-m)\cos^2\theta \bigr\rbrack
+2E\abspr \cos\theta \;. 
\label{alphastandard}
\ear

\subsection{Helicity basis}

Due to our convention of the incoming particle moving in the $z$ direction,
changing to the helicity basis leaves $r^\mu$ unchanged. The vectors $z^\mu$ and $d^\mu$
become simpler:
\bear
z^\mu &=& \frac{1}{m} (\abspr,E'\sin\theta,0,E'\cos\theta)
\label{app-zhel}\\
d^\mu &=& \frac{1}{2}(0,\cos\theta,-i,-\sin\theta) \;.
\label{app-dhel}
\ear
The coefficients also simplify,
\bear
4m^2 \alpha_1 &=& (EE'-\absp\abspr +m^2) (1+\cos\theta) 
\nonumber\\
4m^2 \alpha_2 &=& - m(E+E')\sin\theta
\nonumber\\ 
4m^2\alpha_3 &=&  - m(\absp+\abspr)\sin\theta
\nonumber\\
4m^2\alpha_4 &=& 
(E\abspr -E'\absp)(1+\cos\theta) \;.
\label{app-alphahelicity}
\ear
Note that in the case of forward scattering, $\theta=0$, both bases coincide. If the scattering is
forward and elastic, $\absp=\abspr$, the
coefficients become trivial, $\alpha_1=1, \alpha_2=\alpha_3=\alpha_4=0$. 

\section{Recursion relations for kernel coefficients}
\label{app-recursion}  
Various relations between the coefficients, $A_{N}$, $B_{N\mu\nu}$ and $C_{N}$ have been presented in the main text. Here we follow a different path, taking advantage of the definition of the kernel as the inverse of a Klein-Gordon-like operator, to derive some recursive identities that hold between coefficients for different numbers of photons. These could be useful if one had already determined the coefficients up to a given $N$, since they can be used to produce the coefficients at order $N+1$ and higher.

\subsection*{Scalar QED}
Beginning with scalar QED the full scalar propagator, $D^{x'x}$, is the inverse of the Klein-Gordon operator in position space so that it satisfies the defining equation:
\begin{equation}
	(-D^{\prime 2} + m^2) D^{x'x} = \langle x' \vert x \rangle = \delta^{D}(x' - x)\, .
	\label{eqDInverse}
\end{equation}
Our path integral representation of the $N$-photon dressed propagator, (I.2.1) and (\ref{linemastercov}) is an inverse for this operator projected onto the subspace multi-linear in photon polarisations. Thus, writing the propagator as a sum over photon numbers,
\begin{equation}
	D = D_{0} + D_{1} + \ldots\,, 
\end{equation}
where the term $D_{N}$ is taken multi-linear in $\varepsilon_{1}\ldots \varepsilon_{N}$, the defining equation (\ref{eqDInverse}) in momentum space becomes
\begin{equation}
	(p^{\prime 2} + m^{2})D^{p\prime p} - e\sum_{i}\varepsilon_{i} \cdot (2p^{\prime} + k_{i})D^{p^{\prime} + k_{i}, p} + e^{2}\sum_{i, j} \varepsilon_{i}\cdot \varepsilon_{j} D^{p^{\prime} + k_{i} + k_{j}, p}\Big\vert_{\textrm{multi}} = (2\pi)^{D} \delta^{D}(p^{\prime} + p)\,,
	\label{DRecurse}
\end{equation}
where we truncate at some multi-linear order. Defining the amputated propagator via
\begin{equation}
	D_{N} := \frac{\widehat{D}_{N}}{(p^{\prime 2} + m^{2})(p^{2} + m^{2})}
\end{equation}
we extract the $\mathcal{O}(\varepsilon_{1}\ldots \varepsilon_{N})$ contribution to find a recursion relation,
\begin{align}
	\widehat{D}^{p^{\prime} p}_{N} &= (p^{2} + m^{2})\Big[e\sum_{i = 1}^{N}\varepsilon_{i}\cdot (2p^{\prime} + k_{i}) D_{N-1}^{p^{\prime} + k_{i}, p}(\varepsilon_{1}, k_{1}; \ldots; \hat{\varepsilon_{i}}, \hat{k}_{i};\ldots; \varepsilon_{N}, k_{N}) \nonumber \\
	&- e^{2} \sum_{i \neq j}^{N} \varepsilon_{i}\cdot \varepsilon_{j} D_{N-2}^{p^{\prime} + k_{i}+ k_{j}, p}(\varepsilon_{1}, k_{1}; \ldots ; \hat{\varepsilon_{i}}, \hat{k}_{i};\ldots; \hat{\varepsilon_{j}}, \hat{k}_{i};\ldots; \varepsilon_{N}, k_{N}) \Big]\, ,
	\label{eqDRecursion}
\end{align}
where as usual the hat indicates that the variable is removed. The terms in brackets represent the addition of extra (scalar QED) vertices glued on to a sum of $(N-1)$- and $(N-2)$-photon amplitudes to ``promote'' them to an $N$-photon amplitude.

In the case of on-shell photons, we can simplify the recursion relation by replacing the polarisation vectors $\varepsilon_{i\,\mu}$ according to
\begin{equation}
\varepsilon_{i} \longrightarrow \dfrac{p' \cdot f_i}{p' \cdot k_i },
\end{equation}
in the recursion relation, which then turns into
\begin{align}
\hspace{-2em}	\widehat{D}^{p^{\prime} p}_{N} &= e^{2} (p^{2} + m^{2}) \,  \sum_{i \neq j}^{N} 
\dfrac{p' \cdot f_i \cdot f_j \cdot p'}{p' \cdot k_i p' \cdot k_j}	\, D_{N-2}^{p^{\prime} + k_{i}+ k_{j}, p}(\varepsilon_{1}, k_{1}; \ldots ; \hat{\varepsilon_{i}}, \hat{k}_{i};\ldots; \hat{\varepsilon_{j}}, \hat{k}_{i};\ldots; \varepsilon_{N}, k_{N}) .
\end{align}
So if we use the $R$-representation of the amplitude, (\ref{Master_es_onshell}), then we produce the $N$-photon propagator already written in a manifestly gauge invariant way. 
\subsection*{Spinor QED}
For the spinor case we proceed analogously, but with respect to the kernel $K$, (\ref{defK}) which is a Green function for the Klein-Gordon plus spin coupling operator,
\begin{equation}
	(-D^{\prime 2} + m^2 + \frac{ie}{2}\sigma^{\mu\nu}F_{\mu \nu}) K^{x'x} =  \delta^{D}(x' - x)\, .
\end{equation}
Going to momentum space and projecting onto terms multi-linear in photon polarisation leads to the requirement that
\begin{align}
	(p^{\prime 2} + m^{2})K^{p\prime p} &- e\sum_{i}\varepsilon_{i} \cdot (2p^{\prime} + k_{i})K^{p^{\prime} + k_{i}, p} - \frac{e}{2}\sum_{i}\sigma^{\mu \nu}f_{i\, \mu \nu}K^{p^{\prime} + k_{i}, p} \nonumber \\
	&+ e^{2}\sum_{i, j} \varepsilon_{i}\cdot \varepsilon_{j} K^{p^{\prime} + k_{i} + k_{j}, p}\Big\vert_{\textrm{multi}} = (2\pi)^{D} \delta^{D}(p^{\prime} + p)\,.
	\label{KRecurse}
\end{align}
If we now also use the identification (\ref{defcalK})
\begin{equation}
	K_{N}^{p^{\prime} p} \equiv (-ie)^{N} \frac{\mathfrak{K}_{N}^{p^{\prime} p}}{(p^{\prime 2} + m^{2})(p^{2} + m^{2})}\, ,
\end{equation}
then we can solve the relation for the $\mathfrak{K}_{N}$, for $N \geqslant 1$:
\begin{align}
	\mathfrak{K}_{N}^{p^{\prime}p} &= i  \sum_{i} \frac{\varepsilon_{i} \cdot (2p^{\prime} + k_{i})}{(p^{\prime} + k_{i})^{2} + m^{2}} \mathfrak{K}_{N-1}^{p^{\prime} + k_{i}, p} + \frac{i}{2}\sum_{i} \frac{\sigma^{\mu \nu} f_{i\, \mu \nu}}{(p^{\prime} + k_{i})^{2} + m^{2}} \mathfrak{K}_{N-1}^{p^{\prime} + k_{i}, p}\nonumber \\
	&+ \sum_{i \neq j} \frac{\varepsilon_{i} \cdot \varepsilon_{j}}{(p^{\prime} + k_{i} + k_{j})^{2} + m^{2}} \mathfrak{K}_{N-2}^{p^{\prime} + k_{i} + k_{j}, p}\, .
\end{align}
This equation is analogous to (\ref{eqDRecursion}), but with the inclusion of the additional spin vertex present in the second order Feynman rules. Now we decompose the matrix structure as in (\ref{defABC})
and find three recursion relations between the coefficients. Firstly for $A$,
\begin{align}
	A_{N}^{p^{\prime}p} &= i  \sum_{i} \Big[\frac{\varepsilon_{i}\cdot (2p^{\prime} + k_{i})}{(p^{\prime} + k_{i})^{2} + m^{2}} A_{N-1}^{p^{\prime} + k_{i}, p}(\hat{\varepsilon}_{i}, \hat{k}_{i}) + \frac{f_{i\, \mu \nu} }{(p^{\prime} + k_{i})^{2} + m^{2}}B_{N-1, \nu \mu}^{p^{\prime} + k_{i}, p}(\hat{\varepsilon}_{i}, \hat{k}_{i}) \Big] \nonumber\\
	&+  \sum_{i \neq j}\frac{\varepsilon_{i} \cdot \varepsilon_{j}}{(p^{\prime} + k_{i} + k_{j})^{2} + m^{2}} A_{N-2}^{p^{\prime} + k_{i} + k_{j}, p}(\hat{\varepsilon}_{i}, \hat{k}_{i}, \hat{\varepsilon}_{j}, \hat{k}_{j})\, ,
\end{align}
then for $B$:
\begin{align}
	B_{N\alpha\beta}^{p^{\prime} p} &=  i  \sum_{i} \Big[\frac{\varepsilon_{i}\cdot (2p^{\prime} + k_{i})}{(p^{\prime} + k_{i})^{2} + m^{2}} B_{N-1, \alpha \beta}^{p^{\prime} + k_{i}, p}(\hat{\varepsilon}_{i}, \hat{k}_{i}) -2 \frac{f_{i\, \mu \beta} }{(p^{\prime} + k_{i})^{2} + m^{2}}B_{N-1}^{p^{\prime} + k_{i}, p, \mu}{}_{\alpha}(\hat{\varepsilon}_{i}, \hat{k}_{i}) \Big] \nonumber \\
	&+\frac{i }{2} \sum_{i} \Big[  \frac{f_{i\, \alpha \beta} }{(p^{\prime} + k_{i})^{2} + m^{2}}A_{N-1}^{p^{\prime} + k_{i}, p} (\hat{\varepsilon}_{i}, \hat{k}_{i}) +  \frac{ {\widetilde f}_{i}^{\alpha \beta} }{(p^{\prime} + k_{i})^{2} + m^{2}}C_{N-1}^{p^{\prime} + k_{i}, p} (\hat{\varepsilon}_{i}, \hat{k}_{i}) \Big]\nonumber \\
	&+  \sum_{i \neq j}\frac{\varepsilon_{i} \cdot \varepsilon_{j}}{(p^{\prime} + k_{i} + k_{j})^{2} + m^{2}} B_{N-2, \alpha \beta}^{p^{\prime} + k_{i} + k_{j}, p}(\hat{\varepsilon}_{i}, \hat{k}_{i}, \hat{\varepsilon}_{j}, \hat{k}_{j}) \, ,
\end{align}
and finally for $C$,
\begin{align}
	C_{N}^{p^{\prime} p} &=  \frac{i  }{2}\sum_{i} \Big[2 \frac{\varepsilon_{i}\cdot (2p^{\prime} + k_{i})}{(p^{\prime} + k_{i})^{2} + m^{2}} C_{N-1}^{p^{\prime} + k_{i}, p}(\hat{\varepsilon}_{i}, \hat{k}_{i}) + \frac{\epsilon^{\mu \nu \rho \sigma} f_{i\, \mu \nu} }{(p^{\prime} + k_{i})^{2} + m^{2}}B_{N-1, \rho \sigma}^{p^{\prime} + k_{i}, p}(\hat{\varepsilon}_{i}, \hat{k}_{i}) \Big] \nonumber \\
	&+  \sum_{i \neq j}\frac{\varepsilon_{i} \cdot \varepsilon_{j}}{(p^{\prime} + k_{i} + k_{j})^{2} + m^{2}} C_{N-2}^{p^{\prime} + k_{i} + k_{j}, p}(\hat{\varepsilon}_{i}, \hat{k}_{i}, \hat{\varepsilon}_{j}, \hat{k}_{j}) \, .
\end{align}
These can be used to convert the $N=2$ coefficients presented in the main text into the coefficients for $N = 3$ and greater numbers of photons.

There are two simplification that can be made on-shell. Firstly, the relations between coefficients, (\ref{idAB1}) and (\ref{idAB2}), allow for the right hand sides of the recursion relations to be written entirely in terms of the $B$ and $\widetilde{B}$. Secondly, since $A_N, B_{N\,\alpha\beta}, C_N$ are transversal on-shell, we can again substitute $\varepsilon_{i} \rightarrow \dfrac{p' \cdot f_i}{p' \cdot k_i }$,
to rewrite the relations in terms of the field strength tensors of the photons. Since all on-shell information is contained in $B_{N\,\alpha\beta}$, it suffices to give this coefficient. For $N=2$, we have 
\begin{align}
	B_{2\alpha\beta}^{p^{\prime} p} &=   \sum_{i} \dfrac{i}{(p^{\prime} + k_{i})^{2} + m^{2}} \Big[  -2  f_{i\, \mu \beta} B_{1}^{p^{\prime} + k_{i}, p, \mu}{}_{\alpha}(\hat{\varepsilon}_{i}, \hat{k}_{i})  
+ \dfrac{1}{2}
 A_{1}^{p^{\prime} + k_{i}, p} (\hat{\varepsilon}_{i}, \hat{k}_{i}) f_{i\, \alpha \beta}    \Big],
\end{align}
with
\begin{eqnarray}
B^{p'p}_{1\, \alpha \beta} &=& \dfrac{i}{2} f_{\alpha\beta}, \qquad \qquad
A^{p'p}_{1} = - \dfrac{i r \cdot f \cdot (p-p') }{r \cdot k}.
\end{eqnarray}
In the general case, ($N > 2$)
\begin{align}
	\hspace{-2.5em}B_{N\alpha\beta}^{p^{\prime} p} &=  i  \sum_{i} \dfrac{1}{(p^{\prime} + k_{i})^{2} + m^{2}} \Big[  
\dfrac{ p \cdot B_{N-1}^{p^{\prime} + k_{i}, p} (\hat{\varepsilon}_{i}, \hat{k}_{i}) \cdot (p'+k_i) }{m^2- p \cdot (p' + k_i) } f_{i\, \alpha \beta} +  \dfrac{ p \cdot \widetilde{B}_{N-1}^{p^{\prime} + k_{i}, p} (\hat{\varepsilon}_{i}, \hat{k}_{i}) \cdot (p'+k_i) }{m^2 + p \cdot (p' + k_i) } {\widetilde f}_{i\alpha \beta} \nonumber \\
& -2  f_{i\, \mu \beta} B_{N-1}^{p^{\prime} + k_{i}, p, \mu}{}_{\alpha}(\hat{\varepsilon}_{i}, \hat{k}_{i})    \Big]-  \sum_{i \neq j}\left(\frac{1}{(p^{\prime} + k_{i} + k_{j})^{2} + m^{2}} \right)
	\left( \dfrac{p' \cdot f_i \cdot f_j \cdot p'}{p' \cdot k_i p' \cdot k_j} \right) B_{N-2, \alpha \beta}^{p^{\prime} + k_{i} + k_{j}, p}(\hat{\varepsilon}_{i}, \hat{k}_{i}, \hat{\varepsilon}_{j}, \hat{k}_{j}) \, .
\end{align}
Now we have a manifestly gauge invariant recursion relation that can be used to determine the coefficients at order $N$ from those of lower order.

\section{Helicity amplitudes in the massless limit} 
\label{app-massless}

In the worldline formalism the massless fermion amplitudes corresponding to \eqref{calMs} can be obtained in two ways. One of them is using the formula \eqref{calMsspinfin}
and computing the limit when $m \rightarrow 0$,

\bear
{\cal M}_{N s^{\prime}s}^{p^{\prime} p}	= \lim_{m \rightarrow 0} \frac{(-ie)^N}{2m}
\bar{u}_{s'}(-p') \mathfrak{K}_N^{p^{\prime}p} u_{s}(p). \label{amp_massless_1}  
\ear
The second way is by setting first $m=0$ in \eqref{calMs}, and then taking the limit when $p^2 \rightarrow 0$. With the aid of equations \eqref{SN} and \eqref{defcalK}, we arrive at
\bear
{\cal M}_{N s^{\prime}s}^{p^{\prime} p}
&=& (-ie)^N \left(
\lim_{p^2 \rightarrow 0} \left.  \left[
\bar{u}_{s'}(-p') \frac{\mathfrak{K}_N^{p^{\prime}p}}{\sqrt{-p^2}} u_{s}(p) \right] \right|_{p'^2=p^2}
 +
i \sum_{i=1}^N \, \bar{u}_{s'}(-p') \,  \slash{\varepsilon}_i
\frac{ \mathfrak{K}_{N-1}^{p^{\prime}+k_i,p} }{(p'+k_i)^{2} } u_s(p) \right), \nonumber \\
\label{am_massless_prev}
\ear
where the spinors in the subleading contribution satisfy the massless on-shell relations in \eqref{onsh}.

The leading contribution in \eqref{am_massless_prev} is equivalent to
\bear
\lim_{m \rightarrow 0}
\frac{(-ie)^N}{m}
\bar{u}_{s'}(-p') \mathfrak{K}_N^{p^{\prime}p} u_{s}(p).
\ear
Then, from \eqref{amp_massless_1} and \eqref{am_massless_prev}, we obtain the second formula to compute the massless amplitude,
\begin{eqnarray}
{\cal M}_{N s^{\prime}s}^{p^{\prime} p}	&=& -e (-ie)^{N-1} \sum_{i=1}^N \, \bar{u}_{s'}(-p') \,  \slash \varepsilon_i
\frac{ \mathfrak{K}_{N-1}^{p^{\prime}+k_i,p} }{(p'+k_i)^{2}} u_s(p) \label{am_massless_2}
\end{eqnarray}
that now comes only from the \textit{subleading} piece that depends upon the kernel with $N-1$ photons.
Let us corroborate that both \eqref{amp_massless_1} and \eqref{am_massless_2} give the correct amplitude for $N=1,2$. According to (I 5.13), (I 5.14), and (I 5.20), we obtain that,
\bear
\mathfrak{K}_0^{p'p} &=& p^2+m^2 = p'^2+m^2 \, ,\label{k_0} \\
\mathfrak{K}_1^{p'p} &=& i\left( \slash \varepsilon {\slash {p}} - {\slash p}' \slash\varepsilon \right) \, ,
\label{k_1} \\
\mathfrak{K}_2^{p'p} &=& 
 \frac{-1}{m^2+(p'+k_1)^2}
 \Bigl\lbrack
- \slash\varepsilon_1(\slash p'+\slash k_1 + m)\slash\varepsilon_2 (\slash p-m) - (\slash p' - m)  \slash\varepsilon_1\slash\varepsilon_2 (\slash p-m)
 \nonumber\\
 && \hspace{100pt}
 +  (\slash p' - m)  \slash\varepsilon_1(\slash p'+\slash k_1 - m)\slash\varepsilon_2 
 \Bigr\rbrack
+ (1\leftrightarrow 2).  \label{k_2}
\ear
Using the formula \eqref{amp_massless_1}, the amplitude for $N=1$ is calculated as,
\begin{eqnarray}
{\cal M}_{1 s^{\prime}s}^{p^{\prime} p}	&=& \lim_{m \rightarrow 0} \frac{e}{2m}
\bar{u}_{s'}(-p') \left( \slash \varepsilon {\slash {p}} - {\slash p}' \slash\varepsilon \right) u_{s}(p) \nonumber \\
&=& \lim_{m \rightarrow 0} \Bigl( -2m \, \frac{e }{2m}
\bar{u}_{s'}(-p')  \slash \varepsilon u_{s}(p) \Bigr) \nonumber \\
&=& -
\bar{u}_{s'}(-p')  \slash \varepsilon u_{s}(p) \, , \label{am_n1_f1}
\end{eqnarray}
where before the limiting process we have used the on-shell relations \eqref{onsh}. The result in \eqref{am_n1_f1} is the same as the one computed from the standard formalism using Feynman diagrams. Similarly for $N=2$, we obtain that 
\begin{eqnarray}
{\cal M}_{2 s^{\prime}s}^{p^{\prime} p}	&=& \lim_{m \rightarrow 0} \frac{e^2}{2m}
\bar{u}_{s'}(-p') \Big(  \frac{1}{m^2+(p'+k_1)^2}
\Bigl\lbrack
- \slash\varepsilon_1(\slash p'+\slash k_1 + m)\slash\varepsilon_2 (\slash p-m) - (\slash p' - m)  \slash\varepsilon_1\slash\varepsilon_2 (\slash p-m)
 \nonumber\\
 && \hspace{100pt}
 +  (\slash p' - m)  \slash\varepsilon_1(\slash p'+\slash k_1 - m)\slash\varepsilon_2 
 \Bigr\rbrack
+ (1\leftrightarrow 2) \Big) u_s(p) \nonumber \\
&=& \lim_{m \rightarrow 0} 2m \frac{e^2}{2m}
\bar{u}_{s'}(-p') \left( \frac{\slash\varepsilon_1(\slash p'+\slash k_1 + m)\slash\varepsilon_2 }{m^2+(p'+k_1)^2} 
+ (1\leftrightarrow 2) \right) u_s(p) \nonumber \\
&=& e^2
\bar{u}_{s'}(-p') \left( \frac{\slash\varepsilon_1(\slash p'+\slash k_1)\slash\varepsilon_2 }{(p'+k_1)^2} 
+ (1\leftrightarrow 2) \right) u_s(p),
\end{eqnarray}
where we have recovered again the result from the standard formalism in the massless case.

Let us compute now the amplitudes with the formula \eqref{am_massless_2}, using \eqref{k_0} and \eqref{k_1} and the massless on-shell relation in \eqref{onsh}. In doing so, we arrive at,
\begin{eqnarray}
{\cal M}_{1 s^{\prime}s}^{p^{\prime} p}	&=& -e \, \bar{u}_{s'}(-p') \,  \slash \varepsilon_{1}
\frac{ (p'+k)^{2} }{(p'+k)^{2}} u_s(p) \nonumber \\
&=& -e \, \bar{u}_{s'}(-p') \,  \slash \varepsilon_1 u_s(p) \, , \\
{\cal M}_{2 s^{\prime}s}^{p^{\prime} p}	&=& -e^2 \, \bar{u}_{s'}(-p') \, \left[  \slash \varepsilon_1
\frac{ \slash \varepsilon_2 {\slash {p}} - ( {\slash p}' + \slash k_1 ) \slash\varepsilon_2 }{(p'+k_1)^{2}} 
+ (1\leftrightarrow 2) \right] u_s(p) \nonumber \\
&=& e^2 \, \bar{u}_{s'}(-p') \, \left[  \slash \varepsilon_1
\frac{ ( {\slash p}' + \slash k_1 )  }{(p'+k_1)^{2}} \slash\varepsilon_2 
+ (1\leftrightarrow 2) \right] u_s(p),
\end{eqnarray}
which give the correct amplitudes.

One of the advantages of working with the formula \eqref{am_massless_2} is that, instead of the coefficients for the kernel $K_N$, we need only those for $K_{N-1}$ which are simpler. In the decomposition \eqref{defABC}, the formula \eqref{am_massless_2} reads as,

\begin{eqnarray}
{\cal M}_{N s's}^{p^{\prime} p}	&=& -e (-ie)^{N-1} \sum_{i=1}^N \, \bar{u}_{\pm}(-p') \,  \slash \varepsilon_i
\frac{ A_{N-1}^{p^{\prime}+k_i,p} + B_{N-1 \, \alpha \beta}^{p^{\prime}+k_i,p} \sigma^{\alpha \beta} - i C_{N-1}^{p^{\prime}+k_i,p} \gamma^5 }{(p'+k_i)^{2}} u_{\pm}(p), \nonumber \\
&=& -e (-ie)^{N-1} \sum_{i=1}^N \, 
\frac{ 1}{(p'+k_i)^{2}} \Big[ \left( A_{N-1}^{p^{\prime}+k_i,p} \mp \, i C_{N-1}^{p^{\prime}+k_i,p} \right)
\bar{u}_{\pm}(-p') \,  \slash \varepsilon_i u_\pm (p) \nonumber \\
&& \hspace{150pt}
+ B_{N-1 \, \alpha \beta}^{p^{\prime}+k_i,p}  
\bar{u}_{\pm}(-p') \,  \slash \varepsilon_i \gamma^\alpha \gamma^\beta u_\pm (p) \Big], 
\label{am_h}
\end{eqnarray}
where in the last line, we have used the antisymmetry of $B_{N-1 \, \alpha \beta}$, and the equation,
\begin{equation}
\gamma^5 u_\pm(p) = \pm u_\pm(p).
\end{equation}

On the other hand in the derivation of the second formula for the massless amplitude, we could have used the reversed identity \eqref{SNreversed}. In this case \eqref{am_h} becomes,
\begin{eqnarray}
{\cal M}_{N s's}^{p^{\prime} p}	
&=& -e (-ie)^{N-1} \sum_{i=1}^N \, 
\frac{ 1}{(p+k_i)^{2}} \Big[ \left( A_{N-1}^{p^{\prime},p+k_i} \mp \, i C_{N-1}^{p^{\prime},p+k_i} \right)
\bar{u}_{\pm}(-p') \,  \slash \varepsilon_i u_\pm (p) \nonumber \\
&& \hspace{150pt}
+ B_{N-1 \, \alpha \beta}^{p^{\prime},p+k_i}  
\bar{u}_{\pm}(-p') \,  \gamma^\alpha \gamma^\beta \slash \varepsilon_i u_\pm (p) \Big].
\label{am_hR}
\end{eqnarray}  

Now, in the spinor helicity formalism the photon polarisation vectors are given by equation \eqref{defpolpm}, and the fermion spinors by \cite{srednicki-book,elvhua-book},
\begin{eqnarray}
u_+(p) &=& | p \rangle, \quad u_-(p) = | p ], \nonumber \\
\bar{u}_+(p) &=& [ p |, \quad \bar{u}_-(p) = \langle p |. \label{spinors_helicity}
\end{eqnarray}

Then, from either \eqref{am_h} or \eqref{am_hR}, the amplitudes when the 
ingoing and the outgoing fermion have different physical helicities vanish since
\begin{eqnarray}
\langle -p' | \; \text{odd} \; \text{number} \; \text{of} \; \gamma \; |p \rangle &=& 0, \nonumber \\ \relax
[ -p' | \; \text{odd} \;  \text{number} \; \text{of} \; \gamma \; |p ] &=& 0 .
\end{eqnarray}
Furthermore, if all photons have the same helicities as the incoming fermion, 
or the opposite helicity of the outgoing one,
the amplitude also vanishes since, choosing the reference momenta of the polarisation vectors to be $-p'$ or $p$, we have that \cite{srednicki-book},
\bear
\slash \varepsilon^{-}(k;p) |p] &=& 0, \quad
[-p'| \, \slash \varepsilon^{-}(k;-p') = 0, \nonumber \\ 
\slash \varepsilon^{+}(k;p) | p \rangle &=& 0, \quad
\langle -p'| \, \slash \varepsilon^{+}(k;-p') = 0. \label{iden_slash_photons}
\ear
These all imply that when all photons have the same helicities, the corresponding tree level amplitude with one incoming and one outgoing massless fermion must vanishes, which agrees with the results of section 8. 

Now, to calculate the non-zero massless amplitudes, we can choose the reference momenta for the photon polarisations such that the terms $\bar{u}_\pm(-p') \, \slash \varepsilon_i \, u_\pm(p)$ in \eqref{am_h}, or equivalently in \eqref{am_hR}, do not contribute to the amplitude. Then, we only need to calculate the terms with the coefficient $B_{N-1 \alpha \beta}$.

For the amplitudes ${\cal M}_{N \;  --}^{ p' p}$
where the incoming and outgoing fermion have negative physical helicity, 
we choose the reference momenta $p$ for the polarisation vectors with negative helicity, and $-p'$ for the polarisation vectors with positive helicities. Then, the amplitude, according to \eqref{am_h}, reads as
\begin{eqnarray}
{\cal M}_{N \;  --}^{ p' p}
&=& 
e (-ie)^{N-1} \sum_{i; \; neg} \, 
\frac{ 2}{ p' \cdot k_i p \cdot k_i} \left( p \cdot f_i^{-} \right)^\alpha
B_{N-1 \, \alpha \beta}^{p^{\prime}+k_i,p}  
\langle -p' | \,  \gamma^\beta \, | p],  \label{s1}
\end{eqnarray}
where we have used the identity
\bear
\varepsilon_i^{\alpha \; -}(k_i,p) = \dfrac{\left( p \cdot f_{i}^- \right)^\alpha}{p \cdot k_i},
\ear
and with the sum running only over photons with negative helicity.

For the amplitudes ${\cal M}_{N \;  ++}^{ p' p}$, we choose the reference momenta to be $p$ for the polarisation vectors with positive helicity, and $-p'$ for the polarisation vectors with negative helicities. Then, the amplitude, according to \eqref{am_hR}, reads as,
\begin{eqnarray}
{\cal M}_{N \;  ++}^{ p' p}
&=& 
-e (-ie)^{N-1} \sum_{i; \; neg} \, 
\frac{ 2}{ p \cdot k_i p' \cdot k_i} \left( p' \cdot f_i^{-} \right)^\alpha
B_{N-1 \, \alpha \beta}^{p^{\prime},p+k_i}  
[ -p' | \,  \gamma^\beta \, | p \rangle. \label{s2}
\end{eqnarray}
%
%

As an example, let us compute the unpolarised-electron cross section for $N=2$. According to \eqref{k_1}, the coefficient $B_{1 \alpha \beta}^{p^{\prime},p}$ is given by,
\begin{equation}
B_{1 \alpha \beta}^{p^{\prime},p} = \dfrac{i}{2} \left( k_{\alpha} \varepsilon_{\beta}
- \varepsilon_{\alpha} k_{\beta}  \right) = \dfrac{i}{2} f_{\alpha \beta}.
\end{equation}
Therefore, from \eqref{s1}, we obtain that
\begin{eqnarray}
{\cal M}_{2 \;  --}^{p'p\; +-}	&=& 
e^2 
\frac{ 1}{ p' \cdot k_2 p \cdot k_2} \left( p \cdot f_2^{-} \cdot
f_1^+  \right)_\beta
\langle -p' | \,  \gamma^\beta \, | p] \nonumber \\
&=& e^2 
\frac{ 1}{2 p' \cdot k_2 p \cdot k_2} [1p]^2 \langle p 2 \rangle \langle  2 (-p') \rangle.
 \label{amT+-+-}
\ear
Similarly, from \eqref{s2} we find
\begin{eqnarray}
{\cal M}_{2 \;  ++}^{p'p\; +-}	&=& 
-e^2 
\frac{ 1}{2 p \cdot k_2 p' \cdot k_2} 
[1(-p')]^2 \langle (-p') 2 \rangle \langle 2 p \rangle. \label{amT-++-}
\end{eqnarray}
Thus the unpolarised-electron cross section turns into
\bear
\langle |{\cal M}_2^{p'p}|^2 \rangle^{+-} &=& 
\dfrac{1}{2} \left( | {\cal M}_{2 \;  - -}^{p'p\; +-}|^2 
+ | {\cal M}_{2 \;  ++}^{p'p \; +-}|^2 \right) \nonumber \\
&=&
- 2 e^4 \left( \dfrac{p'\cdot k_2}{p' \cdot k_1} +  \dfrac{p'\cdot k_1}{p' \cdot k_2} \right).
\ear
This agrees with (\ref{comptonunpol}) in the massless limit.

\end{document}